\pdfoutput=1
\documentclass[preprint]{vldb}

\usepackage[table]{xcolor} 
\usepackage[utf8]{inputenc}
\usepackage[T1]{fontenc}
\usepackage{amsmath}
\usepackage{graphicx}
\usepackage[scaled=0.8]{beramono}
\usepackage{listings}
\usepackage{tikz}
\usepackage{flushend}
\usepackage{listings}
\usepackage{pgfplotstable}
\usepackage[breaklinks]{hyperref}

\lstdefinelanguage{Scala}%
{morekeywords={abstract,%
  case,catch,char,class,%
  def,else,extends,final,finally,for,%
  if,import,implicit,%
  match,module,%
  new,null,%
  object,override,%
  package,private,protected,public,%
  for,public,return,super,%
  this,
  trait,try,type,%
  val,var,%
  with,while,%
  yield%
  },%
  sensitive,%
  morecomment=[l]//,%
  morecomment=[s]{/*}{*/},%
  morestring=[b]",%
  morestring=[b]',%
  showstringspaces=false%
  inputencoding=utf8,
  extendedchars=true,
  moredelim=[is][\overbar]{`}{`},
  moredelim=[is][\sout]{~}{~},
  literate={⇓}{{\raisebox{0.5pt}{$\Downarrow$}}}1 
           {≠}{@}1 
           {↓}{{\raisebox{0.5pt}{$\downarrow$}}}1 
           {∈}{{\raisebox{0.5pt}{$\scriptscriptstyle\in$}}}1 
           {λ}{{{$\lambda$}}}1 
}[keywords,comments,strings]%

\definecolor{dkgreen}{rgb}{0,0.6,0}
\definecolor{gray}{rgb}{0.5,0.5,0.5}
\definecolor{mauve}{rgb}{0.58,0,0.82}
\definecolor{violet}{rgb}{0.53, 0.0, 0.69}
\definecolor{violet2}{rgb}{0.93, 0.51, 0.93}
\definecolor{backcolour}{rgb}{0.95,0.95,0.92}

\lstdefinestyle{myScalastyle}{
}

\lstdefinestyle{myScalastyle2}{
	backgroundcolor=\color{backcolour}
}

\newcommand{\keywordstyle}[1]{\bfseries{#1}}

\lstnewenvironment{listing}{\lstset{language=Scala}}{}

\newenvironment{sitemize}{\vspace{-4pt}
\begin{itemize}
  \setlength{\itemsep}{1pt}
  \setlength{\parskip}{0pt}
  \setlength{\parsep}{0pt}
}{\end{itemize}}

\newcommand{\comment}[1]{}

\newcounter{example}

\usepackage{subcaption}

\tolerance = \pretolerance
\begin{document}


\title{Flare: Native Compilation for Heterogeneous Workloads\\ in Apache Spark}


\numberofauthors{1}

\author{
\alignauthor
Grégory M. Essertel$^1$,
Ruby Y. Tahboub$^1$,
James M. Decker$^1$,\\
Kevin J. Brown$^2$,
Kunle Olukotun$^2$,
Tiark Rompf$^1$\\[4pt]
\affaddr{\textsuperscript{1}Purdue University, \textsuperscript{2}Stanford University}
\email{\{gesserte,rtahboub,decker31,tiark\}@purdue.edu, \{kjbrown,kunle\}@stanford.edu}\\
}

\toappear{[Preprint, March 2017. Copyright by the authors.]}

\maketitle

\lstMakeShortInline[keywordstyle=,%
              flexiblecolumns=false,%
              mathescape=false,%
              basicstyle=\tt]@

\begin{abstract}
The need for modern data analytics to combine relational, procedural, and 
map-reduce-style functional processing is widely recognized. State-of-the-art 
systems like Spark have added SQL front-ends and relational 
query optimization, which promise an increase in expressiveness
and performance. But how good are these extensions at extracting high performance from modern hardware platforms? 

While Spark has made impressive progress, 
we show that for relational workloads, there is still a significant gap 
compared with best-of-breed query engines. And when stepping outside of the relational world, query optimization techniques are ineffective if
large parts of a computation have to be treated as user-defined 
functions (UDFs).

We present Flare: a new back-end for Spark that brings performance
closer to the best SQL engines, without giving up the added expressiveness 
of Spark. We demonstrate order of magnitude speedups both for relational 
workloads such as TPC-H, as well as for a range of machine learning kernels 
that combine relational and iterative functional processing.

Flare achieves these results through (1) compilation to native code, 
(2) replacing parts of the Spark runtime system, and
(3) extending the scope of optimization and code generation to 
large classes of UDFs.

\end{abstract}

\section{Introduction}

\begin{figure*}
  \centering
  \includegraphics[width=6in,height=2.55in]{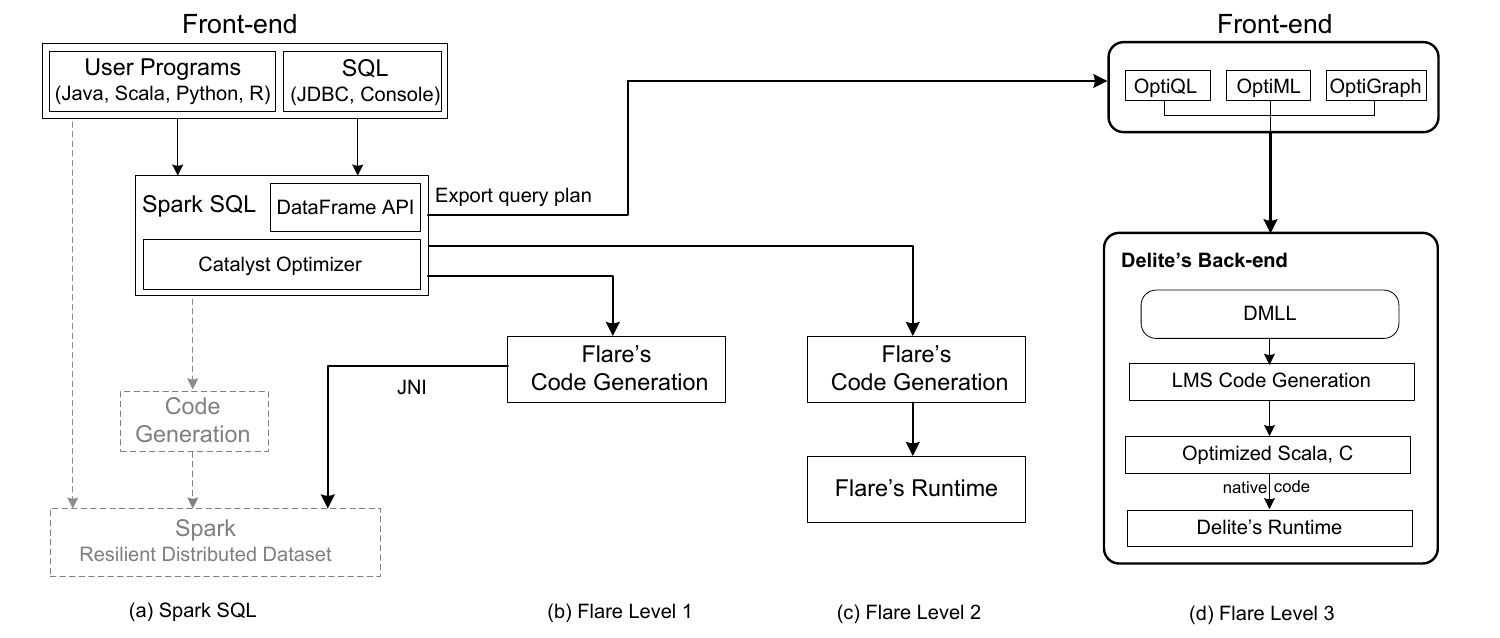}
  \caption{\label{fig:flare}Flare system overview: 
  (a) architecture of Spark and Spark SQL, adapted from 
  \protect\cite{DBLP:conf/sigmod/ArmbrustXLHLBMK15};
  (b) Flare Level 1 replaces Spark's JVM code generation with native 
  code generation, at the granularity of Spark's query stages;
  (c) Flare Level 2 generates code for entire queries, 
  eliminating the RDD layer, and orchestrating parallel execution
  optimized for shared memory architectures;
  (d) Flare Level 3 translates Spark query plans to Delite,
  an established framework for high-performance domain-specific
  languages, which enables optimization of relational queries
  together with UDFs implemented in one of the front-end DSLs.\vspace{-3ex}
  }
\end{figure*}

Modern data analytics applications require a combination of different
programming paradigms, spanning relational, procedural, and map-reduce-style 
functional processing. Shortcomings in both expressiveness and performance
are key reasons why the excitement around early MapReduce tools, heralded as the solution 
to all big data problems, has tapered off. Instead, state-of-the-art systems like 
Spark have added SQL front-ends and APIs which enable relational query 
optimization \cite{DBLP:conf/sigmod/ArmbrustXLHLBMK15}. 

But how good are these extensions? We demonstrate that on standard
relational benchmarks such as TPC-H, Spark SQL still performs at least an 
order of magnitude worse than best-of-breed relational query engines like
HyPer~\cite{DBLP:journals/pvldb/Neumann11}. This is, in part, due to the fact that these relational query engines not only optimize query 
plans aggressively on the relational operator level, but also compile queries 
into native code, thus operating closer to the metal than Spark's current 
Java-based techniques.

While one might argue that this is not a fair comparison due to the added 
expressiveness of Spark and the different nature of these systems, we show 
that it is actually possible to bring the performance of Spark much 
closer to such highly-optimized relational engines without sacrificing this added expressiveness. We present Flare, a 
new back-end for Spark that yields significant speedups by compiling 
entire query plans obtained from Spark's query optimizer to native code, bypassing inefficient abstraction layers of the Spark runtime system.

\vspace{-3pt}
\paragraph*{Heterogeneous Workloads}

While Flare's native code translation alone already provides excellent 
performance for SQL queries and DataFrame operations, performance still suffers 
for heterogeneous workloads, e.g.,\ when executing many small queries 
interleaved with user code, or when combining relational with iterative
functional processing, as is common in machine learning 
pipelines. While the extract, transfer, and load part (ETL) of such pipelines can often be implemented 
as DataFrames, the compute kernels often have to be supplied as 
user-defined functions (UDFs), 
which appear as black boxes to the query optimizer, and thus remain 
as unoptimized library code.

For such workloads, Flare 
takes advantage of Delite \cite{Brown2016GCO,pact11delite,sujeeth2014delite}, 
an existing compiler framework 
for high-performance domain-specific languages (DSLs). 
As an alternative to generating target code in a single step, Flare can map Spark's 
query plans to Delite's intermediate language, DMLL. This enables Spark to 
interface with UDFs written in any of the existing Delite DSLs, which cover 
domains such as machine learning (OptiML \cite{icml11optiml}), graph processing 
(OptiGraph \cite{ecoop13sujeeth}), or mesh-based PDE solvers 
(OptiMesh \cite{ecoop13sujeeth}). All of these DSLs are \emph{embedded} in
Scala and provide APIs comparable to those built on top of Spark. But unlike
Scala code that uses normal Spark APIs, Delite DSL code is amenable to 
optimization and native code generation, including GPU code for
computational kernels \cite{micro2011delite,micro14lee}. We show that combining 
Spark SQL queries and DataFrames with machine learning kernels written in OptiML, 
in particular, results in order of magnitude speedups compared to a standard 
Spark implementation.

\paragraph*{Scale-Up over Scale-Out}

With the implementation of Flare, we revisit some design decisions of 
Spark SQL rooted in the legacy of Spark and earlier systems. We show that 
alternative implementations, which start from different assumptions, 
can provide better performance with the same expressive user-facing API.

One key assumption of MapReduce, Hadoop, and Spark is that of a Google-scale, 
distributed, shared-nothing architecture, focusing primarily on scale-out 
vs scale-up. This makes it easy to scale computation by adding more 
machines, but it also means that each individual machine may not be 
used efficiently \cite{DBLP:conf/hotos/McSherryIM15}.
An immediate consequence is an increase in datacenter
bills, but on a global scale, the effects of scale-out-first 
approaches (a.k.a\ ``nobody ever got fired for running a Hadoop cluster'')
and their inefficient use of energy may be as far-reaching as 
accelerating global warming.

Today, machines with dozens of cores and memory in the TB range are 
readily available, both for rent and for purchase. At the time of writing,
Amazon EC2 instances offer up to 2 TB main memory, with 64 cores and 128 
hardware threads. 
Built-to-order machines at Dell can be configured with up to 12 TB, 96 cores 
and 192 hardware threads. 
NVIDIA advertises their latest 8-GPU system
as a ``supercomputer in a box,'' with compute power equal to hundreds
of conventional servers \cite{nvidiaDGX1}.
With such powerful machines becoming increasingly
commonplace, large clusters are less and less frequently needed. 
Many times, ``big data'' is not \emph{that} big, and often computation is
the bottleneck \cite{DBLP:conf/nsdi/OusterhoutRRSC15}. As such, a small cluster
or even a single large machine is sufficient, 
\emph{if it is used to its full potential}. 

With this scenario as the primary target -- heterogeneous workloads and 
small clusters of powerful machines, potentially with accelerators such 
as GPUs -- Flare prioritizes bare-metal 
performance on all levels of scaling, with the option of bypassing mechanisms such
as fault-tolerance for shared-memory-only execution.
Thus, Flare strengthens Spark's role as a unified, \emph{efficient}, 
big data platform, as opposed to a mere cluster computing fabric.

The overall architecture of Flare, with three possible levels of
integration, is shown in Figure~\ref{fig:flare}. 
The paper first reviews the design of Spark and Spark SQL 
(Section~\ref{sec:spark}) and continues with our main
contributions:
\begin{sitemize}

\item We present Flare: a new accelerator back-end for Spark.
We identify key impediments to performance in Spark 
SQL, particularly related to Spark's Java execution
environment. As an immediate remedy, Flare Level 1 generates 
native code instead of Java for improved performance
(Section~\ref{sec:flare}).

\item We identify further performance issues in Spark SQL,
specifically related to joins in shared-memory environments.
Flare Level 2 drastically reduces these 
overheads by making different starting assumptions
than Spark, focusing on scale-up instead of scale-out, and 
reducing constant factors throughout. In particular,
Flare Level 2 compiles whole queries at once, as opposed to 
individual query stages, which results in an end-to-end
optimized data path
(Section~\ref{sec:flare_level2}).

\item We extend Flare further to accelerate heterogeneous workloads,
consisting of relational queries combined with iterative
machine learning kernels written as user-defined functions.
Flare Level 3 uses the Delite compiler framework to optimize 
relational queries together with such UDFs, by means of
an intermediate language
(Section~\ref{sec:flare-level3}).

\item We evaluate Flare in comparison to Spark both on TPC-H,
reducing the gap to best-of-breed relational query engines, and
on benchmarks involving heterogeneous machine learning workloads.
In both settings, Flare exhibits order-of-magnitude speedups.
Our evaluation spans single-core, multi-core, NUMA, cluster,
and GPU targets
(Section~\ref{sec:flareEval}).

\end{sitemize}
\section{Background on Spark}\label{sec:spark}

Apache Spark~\cite{zaharia10spark,SparkCACM} is today's most popular and most 
widely-used big data framework. The core programming abstraction is an immutable, 
implicitly distributed, collection data type called RDD (Resilient Distributed 
Dataset). RDDs serve as high-level programming interfaces and also transparently
manage fault-tolerance.

Here is a quick example (from \cite{DBLP:conf/sigmod/ArmbrustXLHLBMK15}) 
that counts the number of errors 
in a (potentially distributed) log file:
\begin{lstlisting}[style=myScalastyle]
val lines  = spark.sparkContext.textFile("...")
val errors = lines.filter(s => s.startsWith("ERROR"))
println("Total errors: " + errors.count())
\end{lstlisting}

Spark's RDDs provide a \emph{deferred} API: in the above example, the calls
to @textFile@ and @filter@ merely construct a computation graph. Actual
computation only takes place at the point where @errors.count@ is invoked. 
Sometimes, RDDs are described as \emph{lazily evaluated}. This is somewhat 
misleading, as a second call to @errors.count@ will re-execute the entire 
computation.\footnote{In its original definition, the term ``lazy evaluation''
means that each term is evaluated only when
needed, and \emph{not more than once} \cite{DBLP:conf/popl/HendersonM76}.}
However, RDDs support memoization via explicit calls to @errors.persist()@,
which will mark the data set to be kept in memory for future operations.

\subsection{Spark SQL and the DataFrame API}

The directed acyclic computation graph represented by an RDD 
describes the distributed operations in a coarse-grained way, 
at the granularity of @map@, @filter@, and so on. 
This level of detail is enough to enable
demand-driven computation, scheduling, and fault-tolerance via selective 
re-computation along the ``lineage'' of a result \cite{zaharia10spark}, 
but it does not provide a full view of the computation applied to each
element of a data set. 
For example, in the code snippet above, the argument to @lines.filter@ is a 
normal Scala closure. This makes it easy to integrate RDD code with 
arbitrary external libraries, but it also means that the given closure
needs to be invoked as-is for every element in the data set. 

Thus, the performance of RDDs suffers from two limitations: first, limited 
visibility for analysis and optimizations, especially standard optimizations 
like join re-ordering for relational workloads expressed as RDDs; and second,
interpretive overhead, i.e.\ function calls for each processed tuple. 
Recent Spark versions aim to ameliorate both of these issues with the
introduction of the Spark SQL subsystem~\cite{DBLP:conf/sigmod/ArmbrustXLHLBMK15}.

\subsection{The Power of Multi-Stage APIs and DSLs}

The core addition of Spark SQL is an alternative API based on DataFrames.\footnote{Internally, Spark distinguishes the type {\tt Dataset[T]}, which
provides a typesafe collection API for elements of type {\tt T}, from the
type {\tt DataFrame = Dataset[Row]}, which provides an untyped API for
rows with arbitrary arity and column names. For the purpose of this paper, 
this difference is immaterial; hence, we use the terms Dataset and 
DataFrame interchangeably.}
A DataFrame is conceptually equivalent to a table in a relational database,
i.e.,\ a collection of rows with named columns. However, like RDDs, the DataFrame
API only records operations, but does not compute the result right away. 
We can write the same example as before:
\begin{lstlisting}
val lines  = spark.read.textFile("...")
val errors = lines.filter($"value".startsWith("ERROR"))
println("Total errors: " + errors.count())
\end{lstlisting}
This is just like the RDD API, in that the call to @errors.count@ will
trigger execution. Unlike RDDs, however, DataFrames capture the full
computation/query to be executed. We can obtain the internal 
representation using
\begin{lstlisting}
errors.explain()
\end{lstlisting}
which produces the following output:
\begin{lstlisting}[language=]
== Physical Plan ==
*Filter StartsWith(value#894, ERROR)
+- *Scan text [value#894] 
     Format: ...TextFileFormat@18edbdbb, 
     InputPaths: ..., 
     ReadSchema: struct<value:string>
\end{lstlisting}
From the high-level DataFrame operations, Spark SQL
internally computes a \emph{query plan}, much like
a relational DBMS. Spark SQL optimizes query plans
using its relational query optimizer called Catalyst, 
and may even generate Java code at runtime to accelerate 
parts of the query plan using a component 
named Tungsten (see Section~\ref{sec:catalyst}).

It is hard to overstate the benefits of this kind of
API, which generates a complete program (i.e.\ query) 
representation at runtime. First, it enables various
kinds of optimizations, including classic relational
query optimizations. Second, one can use this
API from multiple front-ends, which exposes
Spark to non-JVM languages such as Python and R,
and the API can also serve as a translation target 
from literal SQL:
\begin{lstlisting}
lines.createOrReplaceTempView("lines")
val errors = spark.sql("select * from lines 
                                 where value like 'ERROR%'")
println("Total errors: " + errors.count())
\end{lstlisting}
Third, one can use the full host language to structure code,
and use small functions that pass DataFrames between them to 
build up a logical plan that is then optimized as a whole.

Of course, this is only true as long as one stays in
the relational world, and does not use UDFs.
As part of our contribution, we will show in Section~\ref{sec:flare-level3} how 
the DataFrame model extends to UDFs in Flare. Flare uses existing
generative programming frameworks \cite{rompf12cacm} to 
make larger classes of expressions available for 
DataFrame-like \emph{multi-stage} programming patterns, where 
a general piece of program code builds up an
intermediate representation (IR) of a more specific
computation at runtime, like DataFrames with query plans.
We argue that such multi-stage APIs, including Spark's DataFrames,
are the real key to success for unified and efficient 
big data platforms, completely independent of RDDs. 
We will show in Section~\ref{sec:sparkPerf} that the 
underlying RDD layer can actually be an impediment 
to performance in Spark.
Flare demonstrates that Spark's DataFrame API can
successfully be supported by a generic, high-performance
DSL compiler framework. This removes the confinement
to relational query optimizations, and enables
optimization of entire data processing pipelines
that combine relational processing with iterative
machine learning kernels that would require 
unoptimized UDFs in plain Spark.

\subsection{Catalyst and Tungsten}
\label{sec:catalyst}

Spark SQL queries are optimized using the Catalyst optimizer that supports both rule-based and cost-based optimizations \cite{DBLP:conf/sigmod/ArmbrustXLHLBMK15}. Starting with a logical query plan tree, Catalyst performs optimizations as tree transformations in order to realize an optimized plan. It is important to note that at the time of writing, Catalyst does not yet perform any kind of join reordering. Instead, it simply joins tables in the order they appear in the @where@ clause. Hence, users need to manually encode a good join in order to avoid big intermediate tables or other bad surprises.

Tungsten is the execution back-end that aims to improve Spark performance by reducing the allocation of objects on the  JVM (Java Virtual Machine) heap, controlling off-heap memory management, employing cache-aware data structures, and generating Java code that is compiled to JVM bytecode at runtime \cite{mastering_apache_Spark2}. These optimizations simultaneously improve the performance of all Spark SQL libraries and DataFrame operations
\cite{SparkCACM}.

\section{Flare: Adding Fuel to the Fire}
\label{sec:flare}

We present Flare: a new back-end for Spark SQL that achieves better performance without changing Spark's user-facing front-end. Flare efficiently compiles 
Spark SQL queries into native code, reduces internal overhead, and brings performance closer to both 
hand-written C and the best-of-breed SQL engines. 

Flare can integrate with Spark at three levels, illustrated in Figure \ref{fig:flare}. Level 1 enables native compilation within Tungsten, at the same level of granularity. Level 2 goes further by compiling whole queries instead of only query \emph{stages}, effectively bypassing Spark's RDD layer and runtime for operations like hash joins in shared-memory environments. Finally, Level 3 goes beyond purely relational workloads by adding another intermediate layer between query plans and generated code. Flare Level 3 employs the Delite compiler framework \cite{micro2011delite} to efficiently compile query pipelines with heterogeneous workloads consisting of user-defined code in multiple DSLs (e.g., SQL and ML).
We will motivate and describe each of these three levels in the
following sections.

\subsection{How Fast is Spark?}
\label{sec:sparkPerf}

Spark performance studies primarily focus on the scale-out performance, e.g., running big data benchmarks \cite{zaharia10spark} on high-end clusters, performing Terabyte sorting \cite{SparkCACM}, etc. 
However, as McSherry, Isard, and Murray have eloquently argued in their 2015 HotOS paper~\cite{DBLP:conf/hotos/McSherryIM15}
and accompanying blog post \cite{cost_blog}, 
big data systems such as Spark tend to scale well, but often just because there is a 
lot of internal overhead.
In particular, McSherry et al.\ demonstrate that a straightforward native implementation of 
the PageRank algorithm \cite{pageRank} 
running on a single laptop can outperform a Spark cluster with 128 cores, using the then-current version.

\begin{figure}
\begin{lstlisting}[style=myScalaStyle]
val tpchq6 = spark.sql("""
  select
    sum(l_extendedprice*l_discount) as revenue
  from
    lineitem
  where
    l_shipdate >= to_date('1994-01-01')
    and l_shipdate < to_date('1995-01-01')
    and l_discount between 0.05 and 0.07
    and l_quantity < 24
""")
\end{lstlisting}
\caption{\label{fig:q6} Query 6 from the TPC-H benchmark in Spark.}
\end{figure}

\vspace{-2ex}
\paragraph*{Laptop vs Cluster}

Inspired by this setup, we are interested in gauging the inherent overheads of Spark and 
Spark SQL in absolute terms. 
\vspace{-1ex}
\begin{quote}
``You can have a second computer once you've shown you know how to use the first one.'' \\– Paul Barham, via \cite{DBLP:conf/hotos/McSherryIM15}
\end{quote}
\vspace{-1ex}

We focus on a simple relational query, which will profit from query optimization and
code generation, and should thus be a best-case scenario for Spark.
We launch a Spark shell configured to use a single worker thread:
\begin{lstlisting}
./bin/spark-shell --master local[1]
\end{lstlisting}

As our benchmark, we pick the simplest query from the industry-standard TPC-H benchmark:
Query 6 (shown in Figure \ref{fig:q6}).
We define the schema of table @lineitem@, provide the source file, and finally register it as a temporary table for Spark SQL (steps not shown). 
For our experiments, we use scale factor 2 (SF2) of the TPC-H data set, which means that table @lineitem@ is stored in a CSV file of about 1.4 GB.
Following the setup by McSherry et al., we run our tests on a fairly standard laptop.\footnote{MacBook Pro Retina 2012, 2.6 GHz Intel Core i7, 16 GB 1600 MHz DDR3, 500 GB SSD,
Spark 2.0, Java HotSpot VM 1.8.0\_112-b16}
We run our query, Q6, straight from the CSV file as input, and record the running time:

\begin{figure}
\begin{lstlisting}[style=myScalaStyle]
// data loading elided ...
for (i = 0; i < size; i++) {
  double l_quantity = l_quantity_col[i];
  double l_extendedprice = l_extendedprice_col[i];
  double l_discount = l_discount_col[i];
  long l_shipdate = l_shipdate_col[i];
  if (l_shipdate >= 19940101L && l_shipdate < 19950101L &&
      l_discount >= 0.05 && l_discount <= 0.07 &&
      l_quantity < 24.0) {
    revenue += l_extendedprice * l_discount;
  }
} ...
\end{lstlisting}
\caption{\label{fig:q6c}Q6 hand-written C code\vspace{-2ex}}
\end{figure}

\begin{figure}
\begin{lstlisting}
Spark SQL             Preload ms      Query ms
    Direct CSV              -           24,400
    Preload CSV          118,062         1,418
Hand-Written C / Flare     
    Preload CSV            2,847            45

\end{lstlisting}
\caption{\label{fig:q6time}Running times for Q6 in Spark, with and without pre-loading, 
and compared to hand-written code and Flare.}
\end{figure}

\begin{lstlisting}
 scala> val q = spark.sql(tpchq6)
 q: org.apache.spark.sql.DataFrame = [revenue: double]
 scala> time(q.show)
 Completed in 24,400 ms
\end{lstlisting}

\noindent 
Clearly, this result of 24 seconds is not the best we can do (see Figure \ref{fig:q6time}). We could convert our data to the columnar Apache Parquet format \cite{Parquet} for increased performance, or we can just preload the data so that subsequent runs are purely in-memory. Since we are mainly interested in the computational part, we opt to preload.
We note in passing that preloading is quite slow (almost 2 min), which may be due to a variety of factors.
Now we can execute our query in-memory, and we get a much better result.
Running the query a few more times yields further speedups, but timings stagnate at around 1s.
The key question now is: how good is this result?

\paragraph*{Hand-Written C}

Since Query 6 is very simple, it is perfectly feasible to write a program in C that performs exactly the same computation: map the input file into memory using the @mmap@ system call, load the data into an in-memory columnar representation, and then execute the main query loop, shown in Figure~\ref{fig:q6c}.
If we compile this C program with @gcc -O3 Q6.c@
and run it, it will take 2.8s in total (including data loading), and \emph{just 45ms} for the actual query computation.
So compared to Spark 2.0, the C program runs 20$\times$ faster!

Is there any good reason \emph{why} the C code needs to be faster than Spark? We believe not, and in fact, running the same query, Q6, accelerated with Flare
yields \emph{exactly the same performance} as the hand-written C code. The same holds for other best-of-breed,
main-memory RDBMSs like HyPer \cite{DBLP:journals/pvldb/Neumann11}, which takes 46.58ms on a comparable machine.

\begin{figure}
\includegraphics[width=\columnwidth]{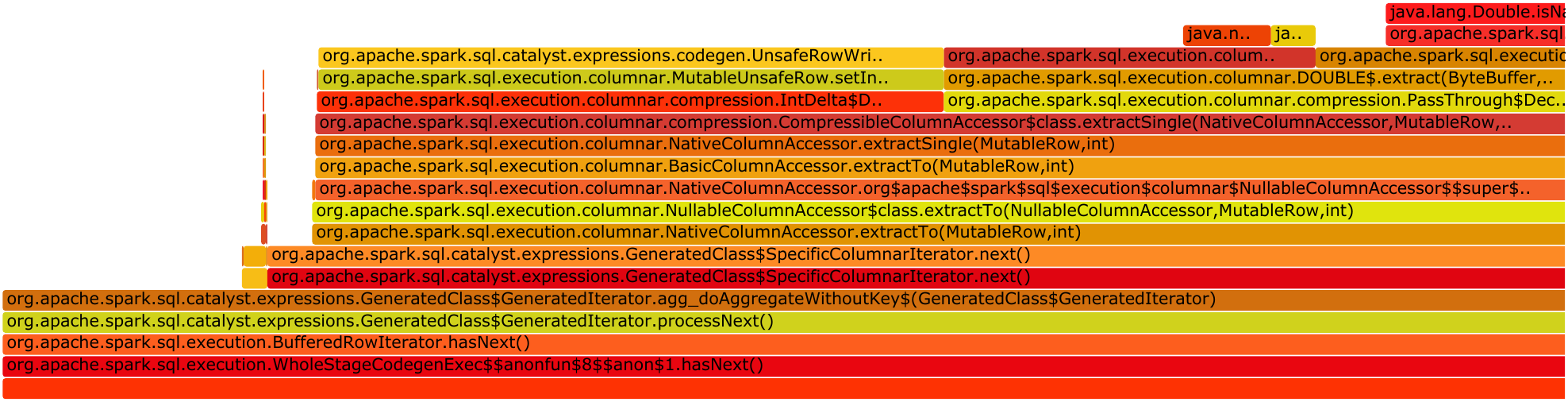}
\vspace{-1ex}
\caption{\label{fig:flamegraph} CPU profile of TPC-H Q6 in Spark SQL, after
pre-loading the {\tt lineitem} table.
80\% of time (the large hump middle to right) is spent accessing and decoding the in-memory data representation.}
\end{figure}

\paragraph*{Identifying the Bottlenecks}

To understand these performance gains, we need to investigate
where Spark SQL spends its time. There are two reasons for the
performance difference, both of which are particularly visible 
in the case of Q6, which has low computational content
and uses only trivial query operators.  First,
we observe that Spark's Tungsten layer generates Java code for this
query. In fact, two pieces of code are generated: one piece for the main 
query loop, the other an iterator to traverse the in-memory data structure.
When looking at the CPU profile (Figure~\ref{fig:flamegraph}), we
can see that 80\% of the execution time is spent in accessing and 
decoding the in-memory data representation, and going back and 
forth between the two pieces of generated code through code paths
that are part of the pre-compiled Spark runtime. Second, even
if we remove this indirection and replace it with a unified
piece of Java code, the performance remains about 30\% lower 
than C, a difference that gets more pronounced for other 
queries where tight low-level control over data structures and
memory management is required.

\subsection{Flare Level 1 Architecture}

Flare Level 1 (Figure \ref{fig:flare}b) adds native compilation within Tungsten. At startup, Flare installs a number of rules into the Catalyst optimization logic. After a query is optimized by Catalyst, these rules will trigger Tungsten to invoke Flare to generate C code for supported operators or combinations of operators. Flare then invokes a C compiler, loads the generated native code into the running JVM through the Java Native Interface (JNI), and Spark's runtime will then execute the generated code as part of the query evaluation. Flare Level 1 \emph{does not} modify any of Spark's other internals, e.g., memory management, data format, RDD layer, etc.  
Hence, Flare Level 1 can serve as a lightweight accelerator that increases performance for certain queries, but not necessarily all, and that does not interfere with Sparks execution model, i.e., transparently supports the standard cluster execution and faul tolerance behavior.
Like Tungsten itself, Flare's query compiler implements Neumann's \cite{DBLP:journals/pvldb/Neumann11} data-centric model, which fuses pipelines of operators that do not need to materialize intermediate results. Code generation in Flare is realized using Lightweight Modular Staging (LMS)~\cite{rompf12cacm}, a generative programming and compiler framework that uses the Scala type system to distinguish normal code from expressions that generate code. In LMS, a special type constructor @Rep[T]@ is used to denote a \emph{staged} expression, which will cause an expression of type @T@ to be emitted in the generated code. 
LMS can be compared to other systems like LLVM \cite{llvm}, but operates on a higher level of abstraction. In the context of query compilation, LMS has been used as part of the LegoBase engine \cite{klonatos2014building}.

\begin{figure}
{\begin{minipage}[b]{0.65\linewidth}
\begin{lstlisting}
Spark
    Sort-merge join          14,937 ms
    Broadcast-hash join       4,775 ms
        of which in exchange  2,232 ms
Flare
    In-memory hash join         136 ms
\end{lstlisting}
\end{minipage}}
\hfill
{\begin{minipage}{0.3\linewidth}
\raggedleft \includegraphics[height=4cm]{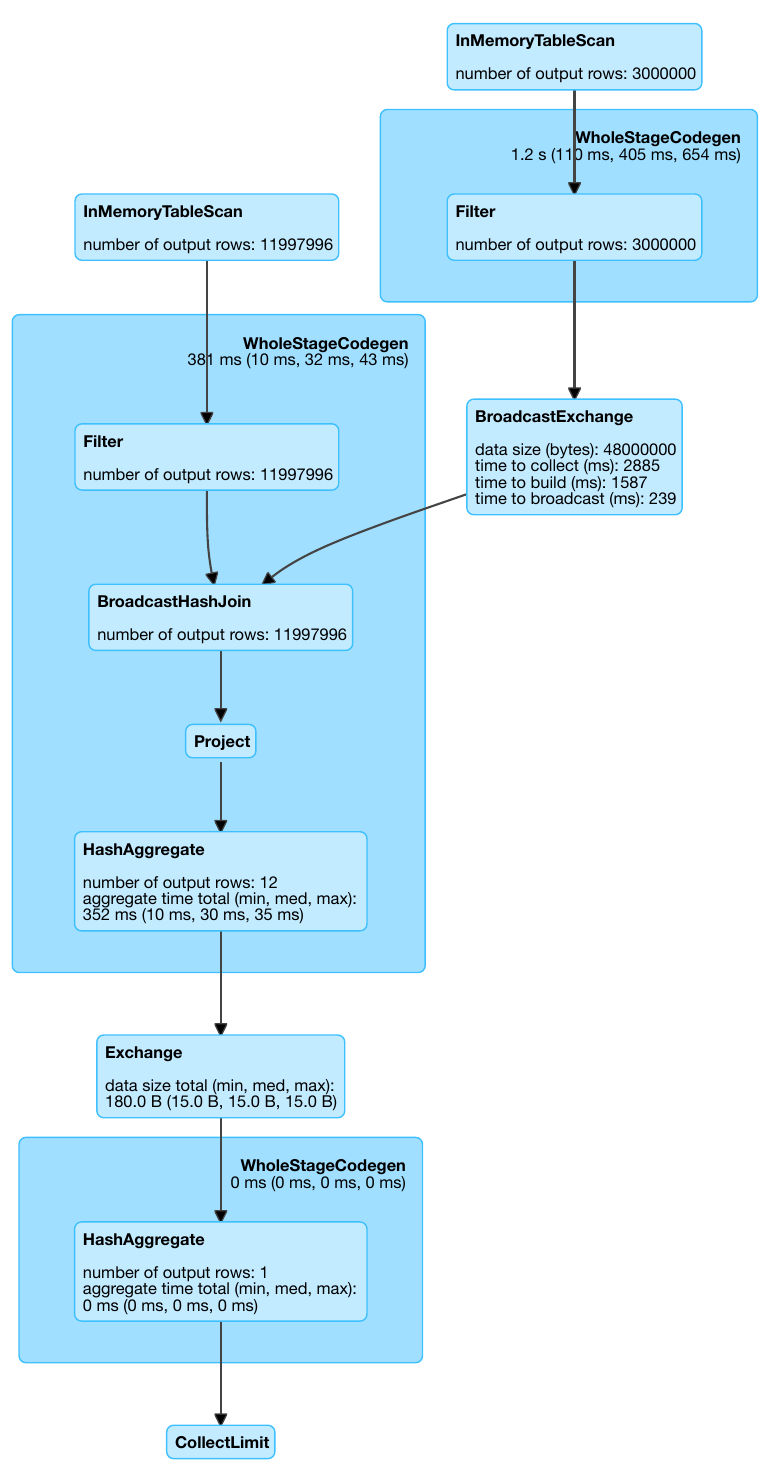}
\end{minipage}}\\
\begin{minipage}{0.7\linewidth}
\vspace{-2cm}
\caption{\label{fig:simple-join}
Join {\tt lineitem} $\bowtie$ {\tt orders}: cost of different operators (left);
Spark's hash join plan (right) shows three
separate code generation regions, which communicate
through Spark's runtime system.}
\end{minipage}
\vspace{-2.4ex}
\end{figure}

\section{Flare Level 2}
\label{sec:flare_level2}
While Flare Level 1 can already provide significant speedups for a number of queries,
it is also constrained by the architecture of the Spark runtime.
Flare Level 2 takes more liberties and revisits key design assumptions of Spark SQL. Spark SQL operates on the legacy of Spark where workloads are assumed to be \emph{Google-scale big}, and the only reasonable way of processing these humongous amounts of data is to scale-out on a large number of distributed, unreliable, shared-nothing machines. Furthermore, Spark SQL inherits Spark's RDD abstraction which encodes lineage to support fault tolerance, which is a necessity in distributed environments of 100s or 1000s of machines. However, many realistic workloads are not \emph{Google-scale}, and, in fact, can be scaled-up on modern big-memory hardware, where faults are statistically highly improbable and fault tolerance can often be supported in hardware, e.g., through RAID arrays, redundant power supplies, or similar facilities. Accelerating Spark SQL's performance \emph{all the way} to the level of best-of-breed modern query engines  such as HyPer~\cite{DBLP:journals/pvldb/Neumann11} requires a \emph{fully} compiled back-end that removes the remaining inefficiencies of legacy Spark.

\paragraph*{More Bottlenecks}
For complex queries, concerns about granularity
of code generation and interfacing with the runtime system
become more pronounced than in our previous example, TPC-H Query 6. 
In fact, queries with joins exhibit some unfortunate
consequences for main-memory execution due to Spark's design
as primarily a cluster computing framework.
Figure~\ref{fig:simple-join} shows timings for a simple 
join query that joins the @lineitem@ and @orders@ tables of 
the TPC-H benchmark. Spark's query optimizer picks an expensive 
sort-merge join by default, which may be the right choice for
distributed or out-of-core execution, but is suboptimal for 
main-memory. With the right flags, it is possible to tune
Spark's Catalyst query planner to prefer a hash join instead,
which is more efficient. But even the hash join operator
follows a broadcast model, with high overhead for the internal 
exchange operator (2.2s of 4.7s), which is present in the 
physical plan even when running on a single core. Looking at
Spark's query plan for this hash join query reveals that
Tungsten has split the query into three code generation 
regions (dark blue areas, right side in Figure~\ref{fig:simple-join}),
which need to communicate through Spark's runtime system. 
It is therefore no surprise that Flare can achieve much faster
query execution by generating tight code for the \emph{entire} 
query.

\begin{figure}
  \centering
  \includegraphics[width=3in,height=1.3in]{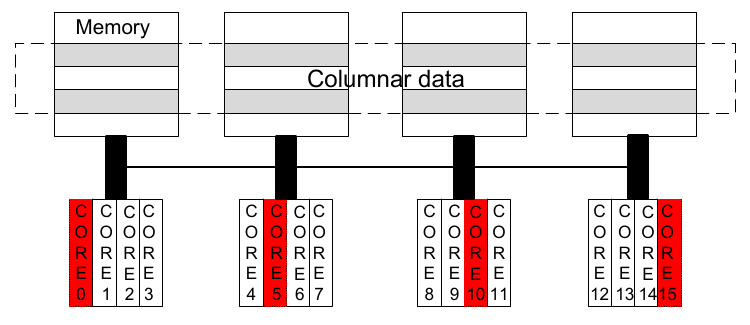}
  \caption{\label{fig:parallel}Thread pinning to specific cores in NUMA.\vspace{-2ex}}
\end{figure}

\subsection{Flare Level 2 Architecture}
The architecture of Flare Level 2 is illustrated in Figure \ref{fig:flare}c. Spark SQL's front-end, DataFrame API, and Catalyst optimizer remain the same. The optimized query plan is exported wholesale from Catalyst to Flare. Flare's compiler iterates recursively over each operator node in the query plan tree and maps it to Flare's internal optimized data structures and query execution logic, represented in LMS. 
After that, LMS performs some light optimizations like common subexpression and dead code elimination, generates C, invokes a C compiler, and Flare launches the resulting binary either inside the JVM, like in Level 1, or as a separate process. This cuts out the rest of Spark and relies solely on Flare's runtime to trigger execution of the generated code. In contrast to Flare Level 1, which operates transparently to the user, Flare Level 2 exposes a dedicated API to let users pick which DataFrames to evaluate through Flare:
\begin{lstlisting}
val df = spark.sql("...") // create DataFrame (SQL or direct)
val fd = flare(df)        // turn it into a FlareDataFrame
fd.show()                 // execute query plan with Flare
\end{lstlisting}

Flare Level 2 does not currently support cluster execution, but it would be possible to extend the Spark runtime with a set of hooks that would delegate to Flare within a single machine, and coordinate multiple cluster nodes and the necessary data exchanges through the existing Spark fabric. Such a setup would be similar to HadoopDB~\cite{hadoopdb}, which combines MapReduce processing between cluster nodes with fast DBMS instances on each node that are able to handle the per-node computation more efficiently than Hadoop.

\subsection{Optimizing Data Loading}
Data loading is an often overlooked factor data processing, and is seldom reported in benchmarks. However, we recognize that data loading from CSV can often be the dominant performance factor for Spark SQL queries. The Apache Parquet~\cite{Parquet} format is an attractive alternative, modeled after Dremel \cite{melnik2010dremel}. As a binary columnar format, it offers opportunities for compression, and queries can load only required columns instead of all data.

In keeping with our running theme, can we do better? While Parquet allows for irrelevant data to be ignored almost entirely, Spark's code to read Parquet files is very generic, resulting in undue overhead. This generality is primarily due to supporting multiple compression and encoding techniques, but there also exists overhead in determining which column iterators are needed. While these sources of overhead seem somewhat unavoidable, in reality they can be resolved by generating specialized code. 
In Flare, we implement compiled CSV and Parquet readers that generate native code specialized to a given schema. As a result, Flare Level 2 can compile data paths end-to-end.

\subsection{Parallel and NUMA Execution}

Query engines realize parallelism either explicitly by implementing special \emph{split} and \emph{merge} operators, or internally by modifying the operator's internal logic to orchestrate parallel execution. Flare Level 2 uses the latter, and realizes parallelism using OpenMP~\cite{OpenMP}. On the architectural level, Flare's operators' implementations take care of splitting their work internally across multiple threads, accumulating final results, etc.
For instance, the parallel @scan@ operator starts a parallel section in its @produce@ method, which sets the number of threads, and invokes @consume@ on the downstream operators in parallel. Join and aggregate operators, in turn, which implement materialization points, implement their @consume@ method in such a way that parallel invocations are possible without conflict, either through per-thread data structures that are merged after the parallel section or through lock-free data structures.

Flare also contains specific optimizations for environments with non-uniform memory access (NUMA), including pinning threads to specific cores and optimizing the memory layout of various data structures to reduce the need for accessing non-local memory. For instance, memory-bound workloads (e.g., TPC-H Q6) perform small amounts of computation, and do not scale-up given a large number of threads on a single CPU socket. Flare's code generation supports such workloads through various data partitioning strategies, in order to maximize local processing and to reduce the need for threads to access non-local memory as illustrated in Figure \ref{fig:parallel} and Section \ref{sec:parallel-scaling}.

\begin{figure}[t!]
  \begin{lstlisting}
1  val tol = 0.001
2  val k = 4
3  def findNearestCluster(x_i: Rep[DenseVector[Double]],
4  mu: Rep[DenseMatrix[Double]]):  Rep[Int] = {
  (mu mapRowsToVector {
            row => dist(x_i, row, SQUARE) }).minIndex}
5  /* Relational ETL */
6  val data = spark.read.csv[Data]("input")
7  val q = data.select(...)
8  val mat = flare(q).toMatrix
9  /* ML DSL with user code (k-Means training loop) */
10 val x = DenseTrainingSet(mat, DenseVector[Double]())
11 val m = x.numSamples
12 val n = x.numFeatures
13 val mu = (0::k, *) { i => x(randomInt(m)) }
14 val newMu = untilconverged_withdiff(mu, tol){ (mu, iter) =>
15     val c = (0::m) { e => findNearestCluster(x(e), mu) }
16     val allWP = (0::m).groupByReduce(...)
17     ...
18 }((x, y) => dist(x, y, SQUARE)) 
19 /* Relational and ML DSLs */
20 val r = spark.sql("""select ... from data
  where class = findNearestCluster(...)
  group by class""")
21 flare(r).show
\end{lstlisting}
  \caption{\label{example:multiDSL}$k$-means clustering in Flare, combining SQL, Spark DataFrames, and OptiML}
\end{figure}

\section{Flare Level 3}
\label{sec:flare-level3}

Many data analytics applications require a combination of different
programming paradigms, i.e., relational, procedural, and map-reduce-style 
functional processing. For example, a machine learning (ML) application might use relational APIs for ETL, and dedicated ML libraries for computations. Spark provides specialized libraries, e.g., ML pipelines, and supports user-defined functions to support domain-specific applications. Unfortunately, Spark's performance falls off a proverbial cliff once DataFrame operations are interleaved with user code. Currently, Spark SQL optimization and code generation treats user code as a black box. Hence, Flare Level 3 focuses on generating efficient code for heterogeneous workloads.

\subsection{User Defined Functions (UDF)}

Spark SQL uses Scala functions which appear as a black box to the optimizer. As mentioned in Section~\ref{sec:flare}, Flare's internal code generation logic is based on a technique called Lightweight Modular Staging (LMS) \cite{rompf12cacm}, which uses a special type constructor @Rep[T]@ to denote \emph{staged} expressions of type @T@, that should become part of the generated code. Extending UDF support to Flare is achieved by injecting @Rep[A] => Rep[B]@ functions into DataFrames in the same way as normal @A => B@ functions in plain Spark. As an example, here is a UDF @sqr@ that squares a given number:
\begin{lstlisting}
// define and register UDF
def sqr(fc: FlareUDFContext) = { import fc._; 
    (y: Rep[Int]) => y * y }
flare.udf.register("sqr", sqr)
// use UDF in query
val df = spark.sql("select ps_availqty from partsupp where
                    sqr(ps_availqty) > 100")
flare(df).show()
\end{lstlisting}
Notice that the definition of @sqr@ uses an additional argument of type @FlareUDFContext@, from which we
import overloaded operators such as @+@, @-@, @*@, etc.,\ to work on @Rep[Int]@ and other @Rep[T]@ types.
The staged function will become part of the code as well, and will be optimized along with the relational operations. This provides benefits for UDFs (general purpose code embedded in queries)
and enables queries to be be optimized with respect to their surrounding code (e.g., queries run within a loop).

\subsection{Heterogeneous Workloads and DSLs}

Flare Level 3 (Figure \ref{fig:flare}d) goes even further and leverages an existing compiler framework called Delite \cite{pact11delite}, which is built on top of LMS, to efficiently compile applications that mix multiple DSLs or ``query languages.''
Delite has a particular focus on parallel and distributed applications, as well as heterogeneous hardware such as GPUs \cite{micro2011delite}. Delite's front-end provides several Scala DSLs, covering various domains, e.g., databases, machine learning, linear algebra, etc. Since these DSLs are \emph{embedded} in Scala, they provide user-facing APIs similar to libraries built on top of Spark.

On the front-end, Flare Level 3 exports a Spark SQL optimized query plan and maps it to Delite's OptiQL (similar to the process explained in Section \ref{sec:flare_level2}).
The core of Delite is built around a first-order functional intermediate language called \emph{Distributed Multi-Loop Language} (DMLL) that models parallel patterns. As the name suggests, DMLL represents parallel patterns as a highly flexible loop abstraction that fuses multiple loops to maximize pipeline and horizontal fusion \cite{Brown2016GCO}. Furthermore, DMLL provides implicitly parallel and nested collection operations, e.g., Collect, Group by-Reduce, as well as  optimizations like loop fusion, loop nesting optimizations, and so on. Finally, DMLL performs analysis and optimizations for multiple hardware targets (e.g. NUMA, clusters, GPU, etc.).

Figure \ref{example:multiDSL} shows code highlights from the popular $k$-means application, which partitions \emph{n} data points into \emph{k} clusters where each data point is assigned to the partition with the nearest mean \cite{berkhin2006survey}. The $k$-means application mixes multiple DSLs, i.e., SQL and OptiML \cite{icml11optiml}, with user code. In lines 6-8, Spark SQL reads data from a file and preprocesses input. Lines 5-16 show the $k$-means processing code using OptiML. Delite provides optimized data types (e.g., @Vectors@) and expressive libraries to assist ML computations. For instance, @mapRowsToVector@, @dist@, and @untilconverged_withdiff@ are ML-specific optimized methods. The final result can be post-processed using SQL as illustrated in lines 20-21.

\begin{figure*}[t]
	\centering
	\includegraphics[width=\textwidth]{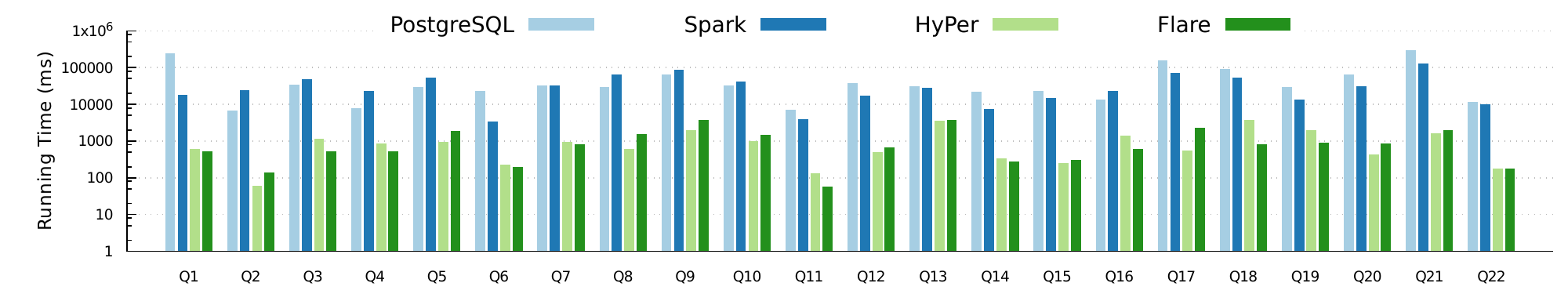}

	{\scriptsize\tt
		\rowcolors{2}{white}{gray!10}
		\setlength{\tabcolsep}{0.25em}
		\begin{tabular}{p{ 0.5in} | r r r r r r r r r r r r r r r r r r r r r r}
			\textbf{SF10} & \textbf{Q1} & \textbf{Q2} & \textbf{Q3} & \textbf{Q4} & \textbf{Q5} & \textbf{Q6} & \textbf{Q7} & \textbf{Q8} & \textbf{Q9} & \textbf{Q10} & \textbf{Q11}	& \textbf{Q12} & \textbf{Q13} & \textbf{Q14} & \textbf{Q15} & \textbf{Q16} & \textbf{Q17} & \textbf{Q18} & \textbf{Q19} & \textbf{Q20} & \textbf{Q21} & \textbf{Q22} \\
			\hline

			\textbf{Postgres}  &
241404 & 6649 & 33721 & 7936 & 30043 & 23358 & 32501 & 29759 & 64224 & 33145 & 7093 & 37880 & 31242 & 22058 & 23133 & 13232 & 155449 & 90949 & 29452 & 65541 & 299178 & 11703 \\

		\textbf{Spark SQL}  &
18219 & 23741 & 47816 & 22630 & 51731 & 3383 & 31770 & 63823 & 88861 & 42216 & 3857 & 17233 & 28489 & 7403 & 14542 & 23371 & 70944 & 53932 & 13085 & 31226 & 128910 & 10030 \\

			\textbf{HyPer}  &
603 & 59 & 1126 & 842 & 941 & 232 & 943 & 616 & 1984 & 967 & 131 & 501 & 3625 & 330 & 253 & 1399 & 563 & 3703 & 1980 & 434 & 1626 & 180\\

		\textbf{Flare} &
			530 & 139 & 532 & 521 & 748
			 & 
			198
			& 830 & 1525 & 3124 & 1436 & 56 & 656 & 3727 & 278 & 302 & 620 & 2343 & 823 & 909 & 870 & 1962 & 177 \\

			\hline

		\end{tabular}
	}
	\caption{\label{fig:sizeSF20}Performance comparison of Postgres, HyPer, Spark SQL, Flare Level 2 in SF10}
\end{figure*} 

\vspace{-2ex}
\section {Experimental Evaluation}
\label{sec:flareEval}

To assess the performance and acceleration potential of Flare in comparison to Spark, 
we present two sets of experiments. The first set focuses on a standard relational 
benchmark; the second set evaluates heterogeneous workloads, consisting of relational
processing combined with machine learning kernels as UDFs. Our experiments span
single-core, multi-core, NUMA, cluster, and GPU targets.

\subsection{Bare-Metal Relational Query Execution}

The first set of experiments focuses on a standard relational workload, and demonstrates 
that the inherent overheads of Spark SQL cause a slowdown of at least 10$\times$
compared to the best available query engines for in-memory execution on a single core.
Our experiments show that Flare Level 2 is able to bridge this gap, accelerating Spark SQL 
to the same level of performance as state-of-the-art query compiler systems, while
retaining the flexibility of Spark's DataFrame API. We also compare
parallel speedups, the effect of NUMA-aware optimization, and evaluate the 
performance benefits of optimized data loading.

\textbf{Environment.} We conducted our experiments on a single NUMA machine with 4 sockets, 12 Xeon E5-4657L cores per socket, and 256GB RAM per socket (1 TB total). The operating system is Ubuntu 14.04.1 LTS. We use Spark 2.0, Scala 2.11, Postgres 9.4, HyPer v0.5-222-g04766a1, and GCC 6.3 with optimization flags @-O3@.

\textbf{Dataset.} We use the standard TPC-H \cite{tpchbib} benchmark with scale factor SF10 for sequential execution, and SF20 and SF100 for parallel execution.

\vspace{-2ex}
\paragraph*{Single-Core Running Time}
In this experiment, we compare the single-core, absolute running time of Flare Level 2 with Postgres, HyPer, and Spark using the TPC-H benchmark with scale factor SF10. In the case of Spark, we use a single executor thread, though the JVM may spawn auxiliary threads to handle GC or the just-in-time compilation.
Postgres and HyPer implement cost-based optimizers that can avoid inefficient query plans, in particular by reordering joins.
While Spark's Catalyst optimizer \cite{DBLP:conf/sigmod/ArmbrustXLHLBMK15} is also cost-based, it does not perform any kind of join re-ordering.
Hence, we match the join ordering of the query plan in Spark SQL and Flare with HyPer's, with a small number of exceptions: in Spark SQL, the original join ordering given in the TPC-H reference outperformed the HyPer plans for Q5, Q9, Q10, and Q11 in Spark SQL, and for Q10 in Flare. For these queries, we kept the original join ordering as is. For Spark SQL, this difference is mainly due to Catalyst picking sort-merge joins over hash joins. It is worth pointing out that HyPer and Postgres plans can use indexes on primary keys, which may give an additional advantage.

Figure \ref{fig:sizeSF20} gives the absolute execution time of Postgres, HyPer, Spark SQL, and Flare for all TPC-H queries. For all systems, data loading time is excluded, i.e., only execution time is reported. In Spark and Flare, we use @persist@ to ensure that the data is loaded from memory. At first glance, the performance of Flare and HyPer lie within the same range, and notably outperform Postgres and Spark in all queries. Similarly, Spark's performance is comparable to Postgres's in most of the queries. Unlike the other systems, Postgres does not compile queries at runtime, and relies on the Volcano model \cite{graefe1994volcano} for query evaluation, which incurs significant overhead. Hence, we can see that Spark's query compilation does not provide a significant advantage over a standard interpreted query engines on most queries.

At a closer look, Flare outperforms Spark SQL in aggregate queries Q1 and Q6 by 34$\times$ and 17$\times$ respectively. We observe that Spark is an order of magnitude slower than Flare in nested queries like those found in Q2. After examining the execution plans of Q2, we found that Catalyst's plan does not detect all patterns that help with avoiding re-computations, e.g., a table which has been previously scanned or sorted.
In join queries, e.g., Q5, Q10, Q14, etc., Flare is faster than Spark SQL by 20$\times$-60$\times$. Likewise, in join variants outer join Q13, semi-join Q21, and anti-join Q22, Flare is faster by 8$\times$, 89$\times$ and 57$\times$ respectively.

The single-core performance gap between Spark SQL and Flare is attributed to the bottlenecks identified in Sections~\ref{sec:flare} and \ref{sec:flare_level2}. First, overhead associated with low-level data access on the JVM. Second, Spark SQL's \emph{distributed-first} strategy that employs costly distributed operators, e.g., sort-merge join and broadcast hash join, even when running on a single core. Third, internal bottlenecks in in-memory processing, the overhead of RDD operations, and communication through Spark's runtime system. By compiling entire queries, instead of isolated query stages, Flare effectively avoids these bottlenecks.

HyPer~\cite{DBLP:journals/pvldb/Neumann11} is a state-of-the-art compiled relational query engine.
A precursory look shows that Flare is faster than HyPer by 10\%-70\% in Q1, Q4-Q6, Q7, and Q14. Moreover, Flare is faster by 2$\times$-4.5$\times$ in Q3, Q11, Q16, Q18, and Q19.
On the other hand, HyPer is faster than Flare by 20\%-60\% in Q9, Q10, Q12, Q15, and Q21. Moreover, HyPer is faster by 2$\times$-4.1$\times$ in Q2, Q8, Q17, and Q20.
This performance gap is, in part, attributed to (1) HyPer's use of specialized operators like GroupJoin~\cite{moerkotte2011groupjoin}, and (2) employing indexes on primary keys as seen in Q2, Q8, etc., whereas Flare (and Spark SQL) currently does not support indexes.
In order to understand the performance gained by using indexes,
we disabled indexing in HyPer and re-ran the benchmarks (detailed results omitted). We observed that Flare outperformed HyPer by 80\% in Q13 and matched the performance of Q12 and Q21. For Q2, Q8-Q10, and Q17, where HyPer outperformed Flare, the performance gap was shrunk by 20\% to 1.2$\times$. Finally, in Q4, Q7, Q11, and Q18 where Flare outperformed HyPer, the performance gap had increased by 10\% to 1.2$\times$.

In summary, while both Flare and HyPer generate native query code at runtime, subtle implementation differences in query evaluation and code generation can result in faster code. For instance. HyPer uses proper decimal precision numbers, whereas Flare follows Spark in using double precision floating point values. which are native to the architecture. Furthermore, HyPer generates LLVM code, whereas Flare generates C code which is then compiled with GCC.

\begin{figure*}
\centering
\includegraphics[width=\linewidth]{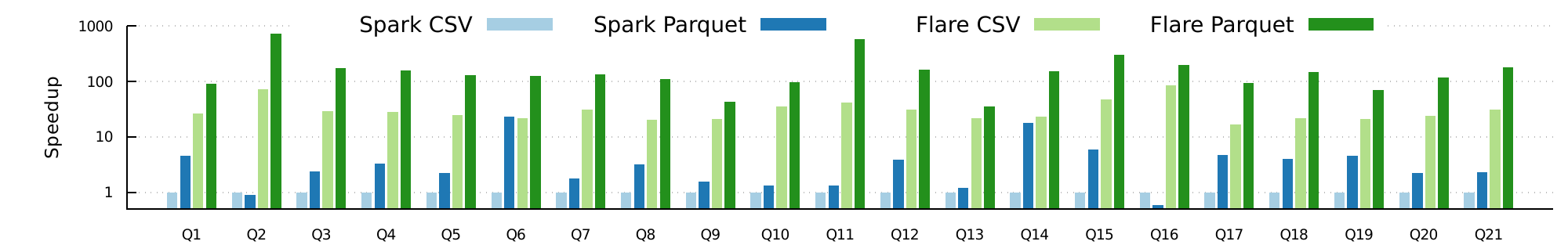}
\caption{\label{fig:streaming-speedup} Speedup for TPC-H SF1 when streaming data from SSD on a single thread.
}
\end{figure*}

\paragraph*{Compilation Time}

We compared the compilation time for each TPCH-H query on Spark and Flare
(detailed results not shown). For Spark, we measured the time to generate the physical plan,
which includes Java code generation and compilation. We do not quantify JVM-internal 
JIT compilation, as this is hard to measure, and code may be recompiled
multiple times. For Flare, we measured C code generation and compilation with GCC.
Both systems spend a similar amount of time on code generation and compilation,
on average 20\% more in Flare. Compilation time depends on the complexity of
the query but is less than 1.5s for all queries,
i.e.,\ well in line with interactive, exploratory, usage.

\paragraph*{Parallel Scaling}
\label{sec:parallel-scaling}

In this experiment, we compare the scalability of Spark SQL and Flare Level 2. The experiment focuses on the absolute performance and the Configuration that Outperforms a Single Thread (COST) metric proposed by McSherry et al.~\cite{DBLP:conf/hotos/McSherryIM15}. We pick four queries that represent aggregate and join variants.

\begin{figure}[t!]
	\includegraphics[width=.49\linewidth]{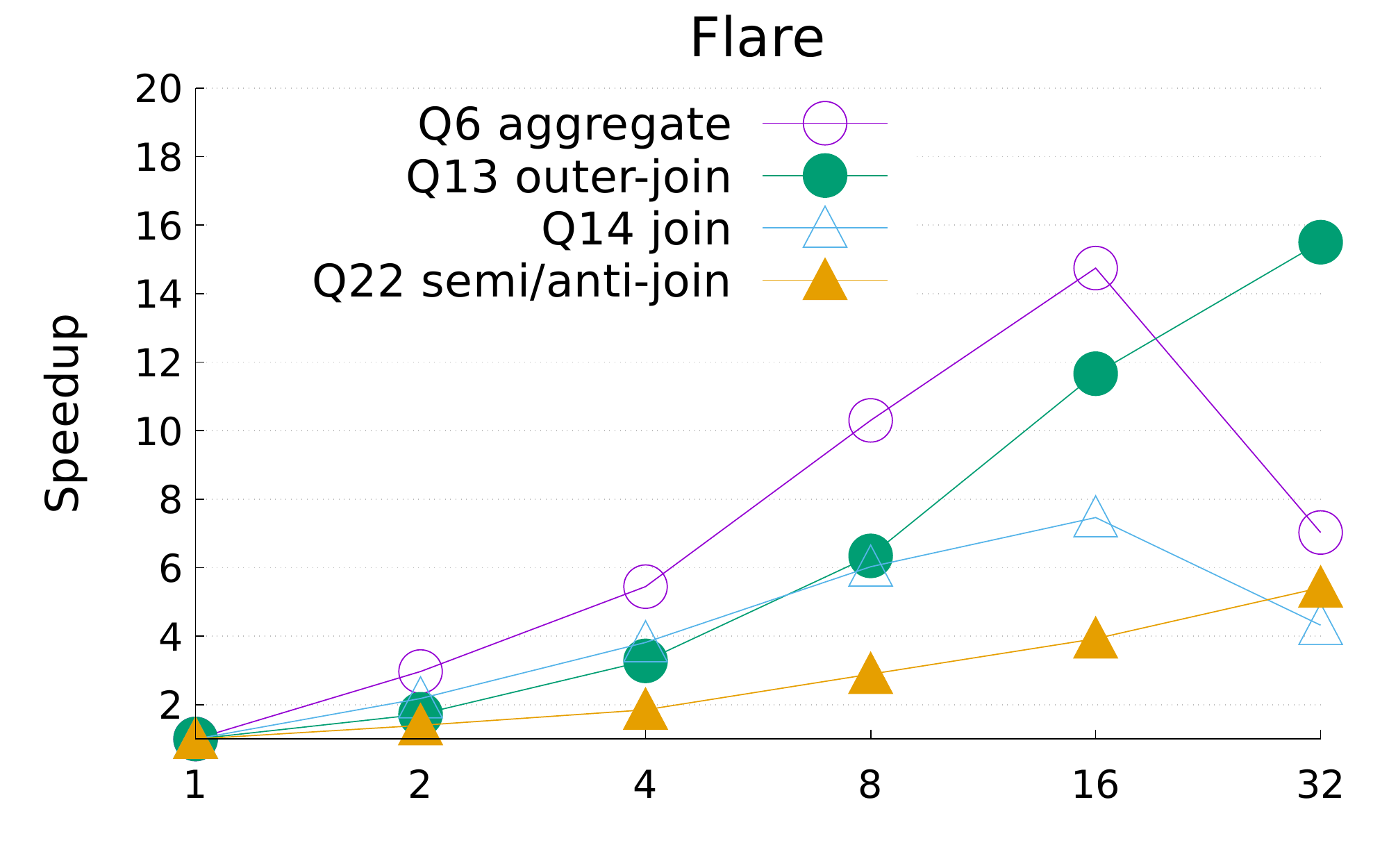}
	\hfill
	\includegraphics[width=.49\linewidth]{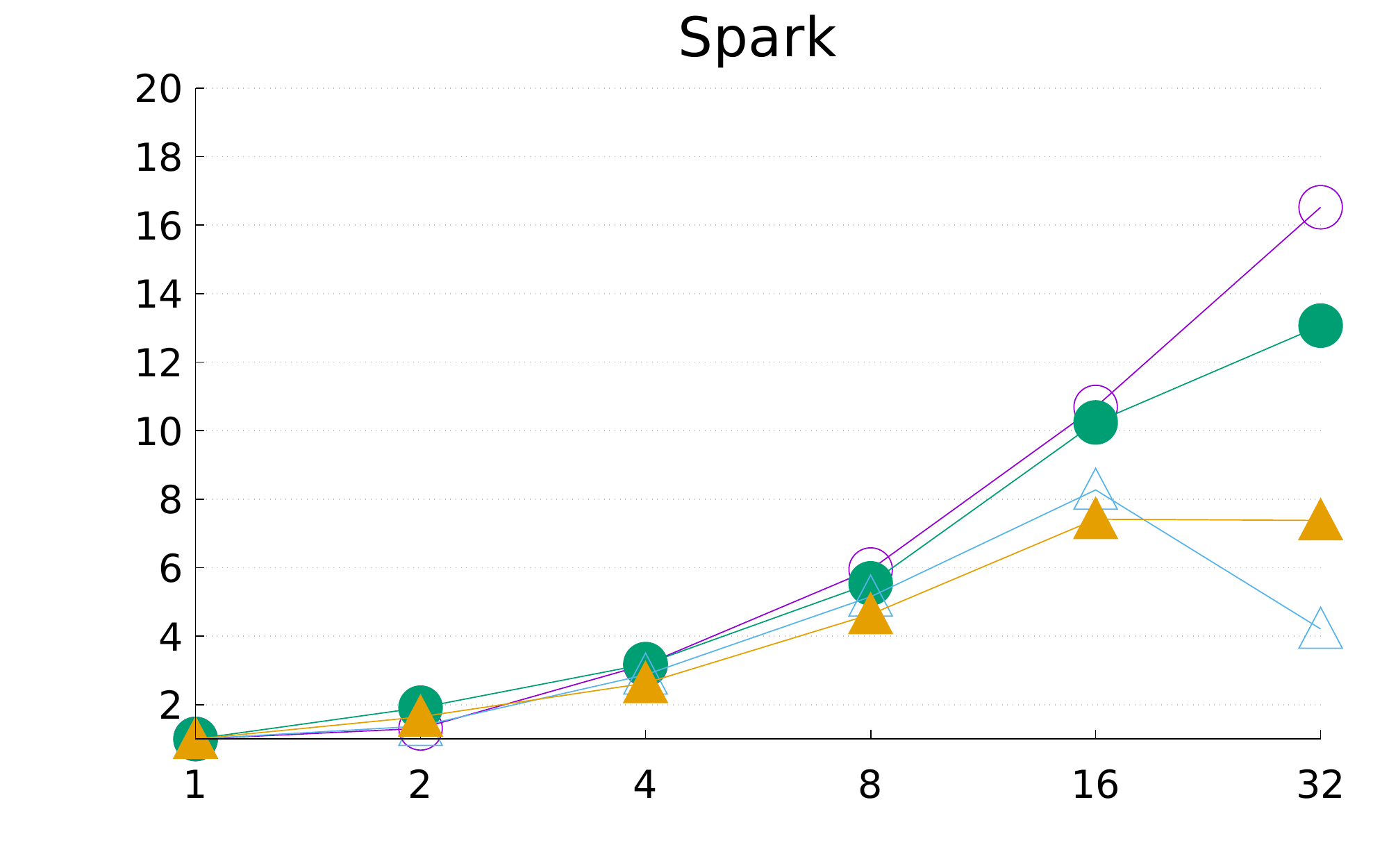}
	\label{img1}
	\\[-4ex]
	\setcounter{subfigure}{0}
	\noindent
	\includegraphics[width=.49\linewidth]{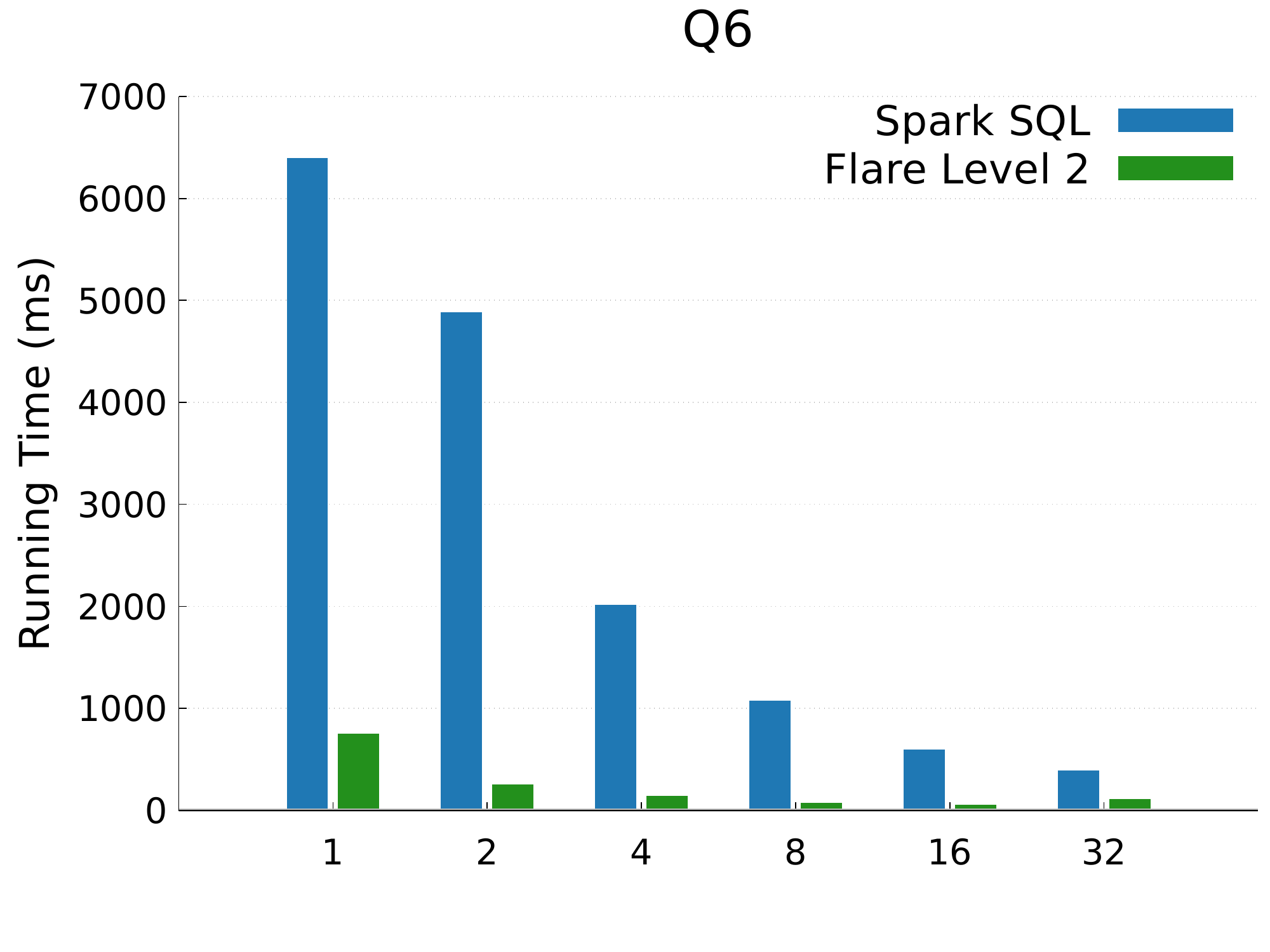}
	\hfill
	\includegraphics[width=.49\linewidth]{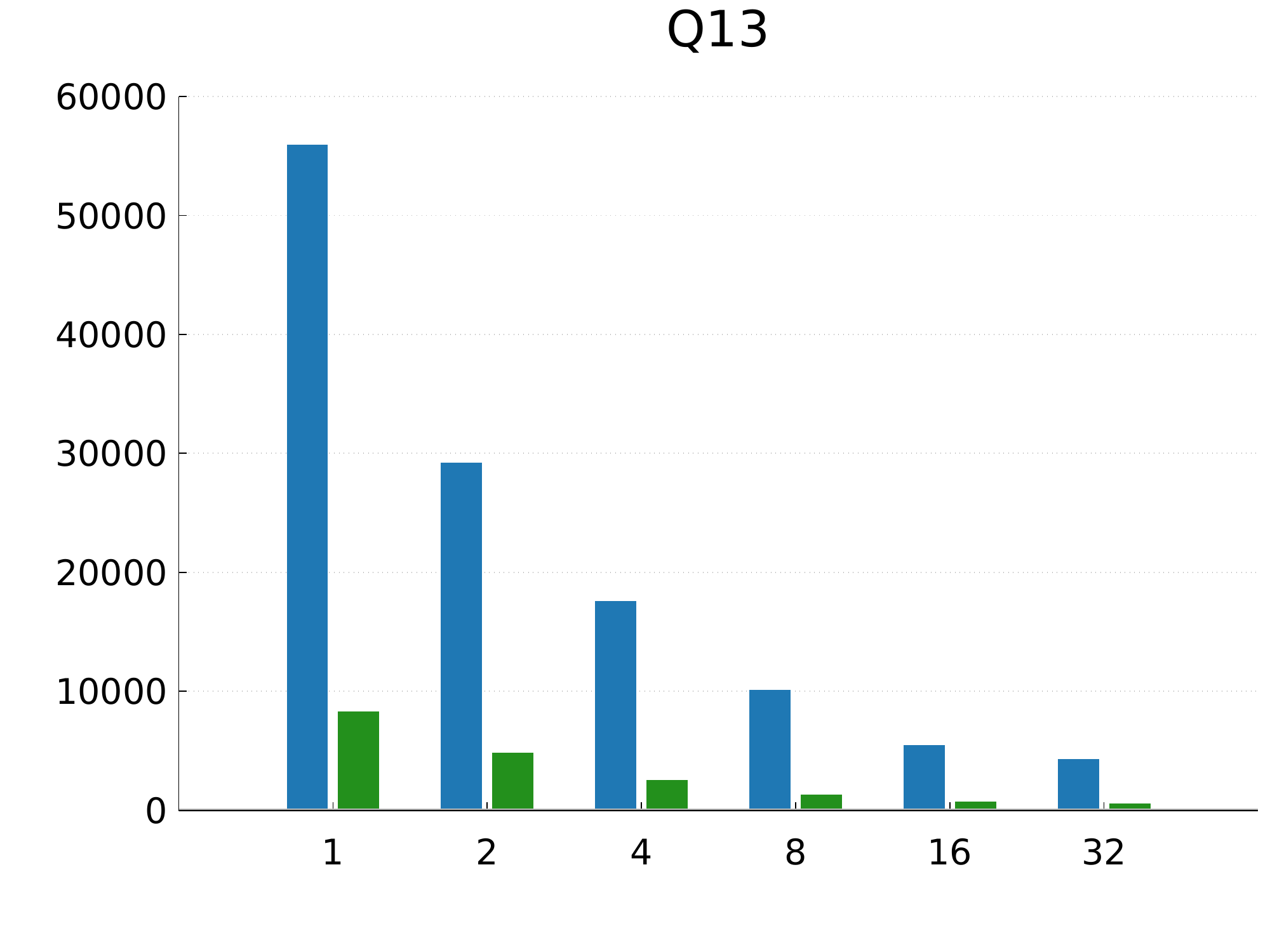}
	\\[-2ex] 
	\includegraphics[width=.49\linewidth]{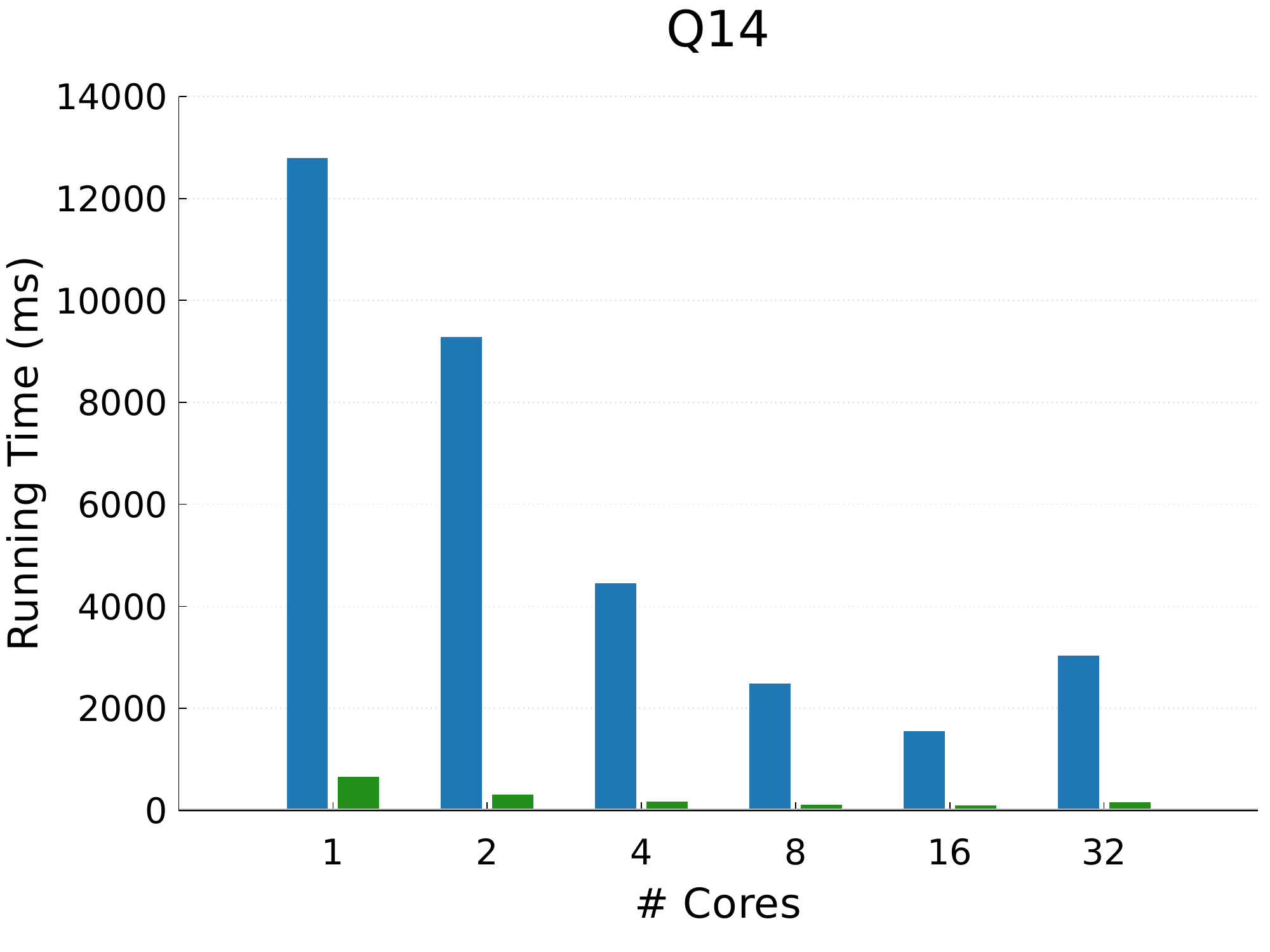}
	\hfill
	\includegraphics[width=.49\linewidth]{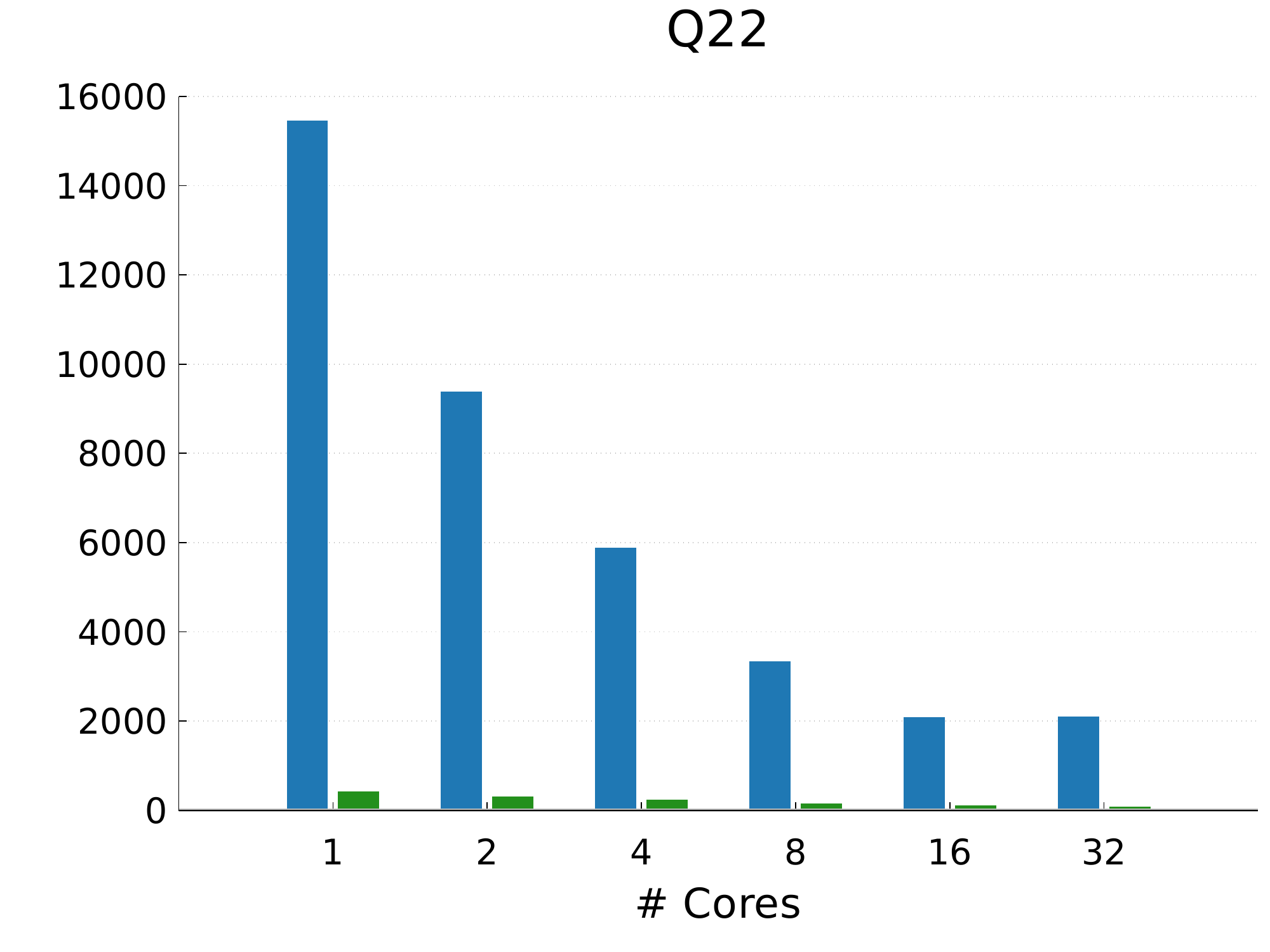}
	
	\caption{\label{fig:speedup20} Scaling-up Flare and Spark SQL in SF20,
		without NUMA optimizations: Spark has good nominal speedups (top), but
		Flare has better absolute running time in all configurations (bottom).
		For both systems, NUMA effects for 32 cores are clearly visible.
	}
	\vspace{-3ex}
\end{figure}

Figure \ref{fig:speedup20} presents speedup numbers for Q6, Q13, Q14, and Q22 when scaled up to 32 cores. At first glance, Spark appears to have good speedups in Q6 and Q13 whereas Flare's Q6 speedup drops for high core counts. However, examining the absolute running times, Flare is faster than Spark SQL by 9$\times$. Furthermore, it takes Spark SQL estimated 12 cores in Q6 to match the performance of Flare's single core. In scaling-up Q13, Flare is consistently faster by 8$\times$ on all cores. Similarly, Flare continues to outperform Spark by a steady 25$\times$ in Q14 and by 20$\times$-80$\times$ in Q22 as the number of cores reaches 32. Notice the COST metric in the last two queries is \emph{infinity}, i.e., there is no Spark configuration that matches Flare's single-core performance.

The seemingly good scaling for Spark reveals that the runtime incurs significant overhead. In particular, we would expect Q6 to become memory-bound as we increase the level of parallelism. In Flare we can directly observe this effect as a sharp drop from 16 to 32 cores. Since our machine has 18 cores per socket, for 32 cores, we start accessing non-local memory (NUMA). The reason Spark scales better is  because the internal overhead, which does not contribute anything to query evaluation, is trivially parallelizable and hides the memory bandwidth effects.
In summary, Flare scales as expected for both of memory and CPU-bound workloads, and reflects the hardware characteristics of the workload, which means that query execution takes good advantage of the available resources -- with the exception of multiple CPU sockets, a problem we address next.

\begin{figure}[t!]
\includegraphics[width=\linewidth]{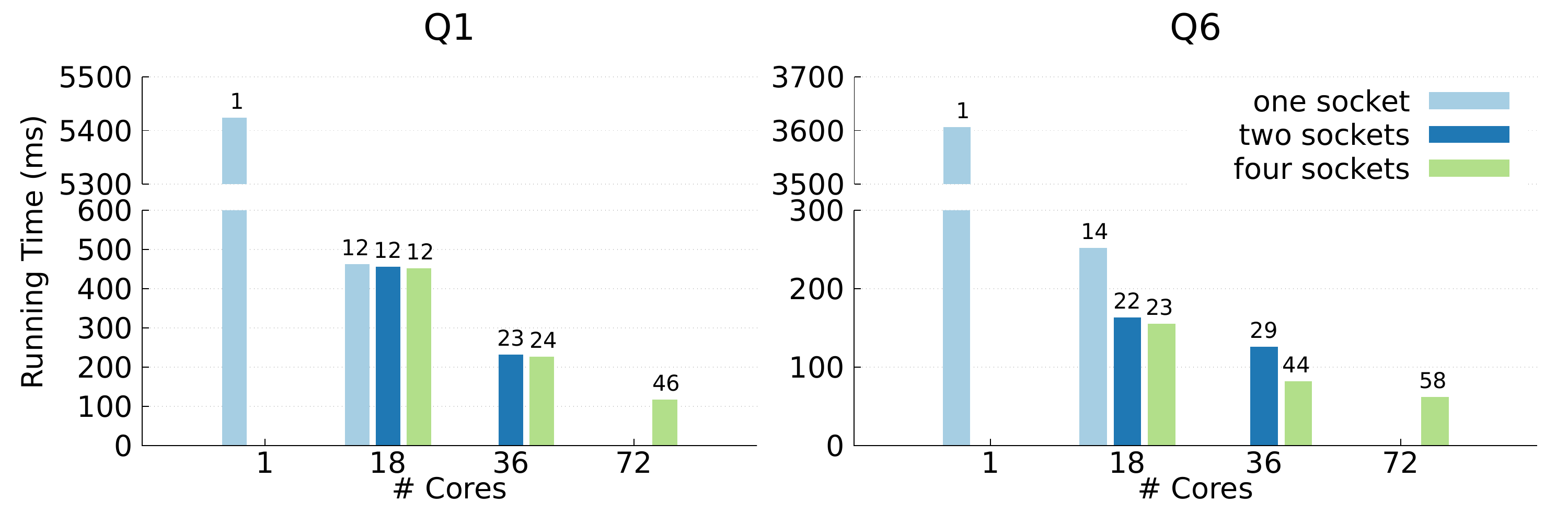}
\caption{\label{fig:tpch-numa-sf100} Scaling-up Flare for SF100 with NUMA optimizations on different configurations: threads pinned to one, two, or four sockets. The speedups relative to a single thread are shown on top of the bars. }
\end{figure}

As a next step, we evaluate NUMA optimizations in Flare and show that these
enable us to scale queries like Q6 to higher core numbers.
In particular, we pin threads to individual cores and lay out memory
such that most accesses are to the local memory region
attached to each socket (Figure~\ref{fig:tpch-numa-sf100}). Q6 performs better when the threads are dispatched on different sockets. This is due to the computation being bounded by the
memory bandwidth. As such, when dividing the threads on multiple sockets, we multiply the available bandwidth proportionally. However, as Q1 is more computation bound, dispatching the threads on different sockets has little effect. For both Q1 and Q6, we see scaling up to the capacity of the machine (in our tests, up to 72 cores). This is seen in a maximum speedup of 46$\times$ and 58$\times$ for Q1 and Q6, respectively.

\begin{table}[t!]
	\centering
	{\scriptsize\tt
		\rowcolors{2}{white}{gray!10}
		\setlength{\tabcolsep}{0.25em}
		\begin{tabular}{|p{ 0.5in} | r | r | r | r | r| r| r|}
			\hline
			\textbf{Table} & \textbf{\#Tuples} & \textbf{Postgres} & \textbf{HyPer} &
			\textbf{Spark} & \textbf{Spark} & \textbf{Flare} & \textbf{Flare} \\
			&  & CSV & CSV & CSV & Parquet & CSV & Parquet \\
			\hline
			\textbf{CUSTOMER} & 1500000 & 7067 & 1102 & 11664 & 9730 & 329 & 266 \\
			\hline
			\textbf{LINEITEM} & 59986052 & 377765 & 49408 & 471207 & 257898 & 11167 & 10668 \\
			\hline
			\textbf{NATION} & 25 & 1 & 8 & 106 & 110 & < 1 & < 1 \\
			\hline
			\textbf{ORDERS} & 15000000 & 60214 & 33195 & 85985 & 54124 & 2028 & 1786 \\
			\hline
			\textbf{PART} & 2000000 & 8807 & 1393 & 11154 & 7601 & 351 & 340 \\
			\hline
			\textbf{PARTSUPP} & 8000000 & 37408 & 5265 & 28748 & 17731 & 1164 & 1010 \\
			\hline
			\textbf{REGION} & 5 & 1 & 8 & 102 & 90 & < 1 & < 1 \\
			\hline
			\textbf{SUPPLIER} & 100000 & 478 & 66 & 616 & 522 & 28 & 16 \\
			\hline
		\end{tabular}
	}
	\vspace{1ex}
	\caption{\label{tbl:loading}Loading time in ms for TPC-H SF10 in Postgres, HyPer, and SparkSQL.\vspace{-4ex}}
\end{table}

\paragraph*{Optimized Data Loading}
\label{sec:optimizedDataLoadingEval}
An often overlooked part of data processing is data loading. Flare contains an optimized
implementation for both CSV files and the columnar Apache Parquet format.\footnote{All Parquet files tested were uncompressed and encoded using Parquet's PLAIN encoding.}
We show loading times for each of the TPC-H tables in Table~\ref{tbl:loading}.

\textbf{Full table read.} From the data in Table~\ref{tbl:loading}, we see that in both Spark and Flare, the Parquet file readers outperform the CSV file readers in most scenarios, despite this being a worst-case scenario for Parquet. Spark's CSV reader was faster in only one case: reading @nation@, a table with only 25 rows. In all other cases, Spark's Parquet reader was 1.33$\times$-1.81$\times$ faster. However, Flare's highly optimized CSV reader operates at a closer level of performance to the Parquet reader, with all tables except @supplier@ having a benefit of less than a 1.25$\times$ speedup by using Parquet.

\textbf{Performing queries.} Figure~\ref{fig:streaming-speedup} shows speedups gained from executing queries without preloading data for both systems. Whereas reading an entire table gives Spark and Flare marginal speedups, reading just the required data gives speedups in the range of 1.22$\times$-22.96$\times$ for Spark (excluding Q2 and Q16, which performed 9\% and 41\% slower, respectively) and 2$\times$-14$\times$ for Flare. Across systems, however, Flare's Parquet reader demonstrated between a 5$\times$-795$\times$ speedup over Spark's, and between 35$\times$-720$\times$ over Spark's CSV reader. While the speedup over Spark lessens slightly in higher scale factors, we found that Flare's Parquet reader consistently performed on average at least one order of magnitude faster across each query, regardless of scale factor.

In nearly every case, reading from a Parquet file in Flare is approximately 2$\times$-4$\times$ slower than in-memory processing (as expected). However, reading from a Parquet file in Spark is rarely significantly slower than in-memory processing. These results show that while reading from Parquet certainly provides performance gains for Spark when compared to reading from CSV, the overall performance bottleneck of Spark surely does not lie in the cost of reading from SSD compared to in-memory processing.

\vspace{-2ex}
\subsection{Heterogeneous Workloads and UDFs}

We now turn our attention to heterogeneous workloads that combine
relational processing with iterative functional computation.
We study a range of machine learning kernels, where the input data
is loaded via DataFrames. The kernel computation is expressed as
a UDF written in the OptiML DSL, which is used as part of a
SQL query.

\begin{figure}
	\centering
	\setcounter{subfigure}{0}
	\vspace{-2ex}
	\noindent
	\includegraphics[width=0.49\linewidth]{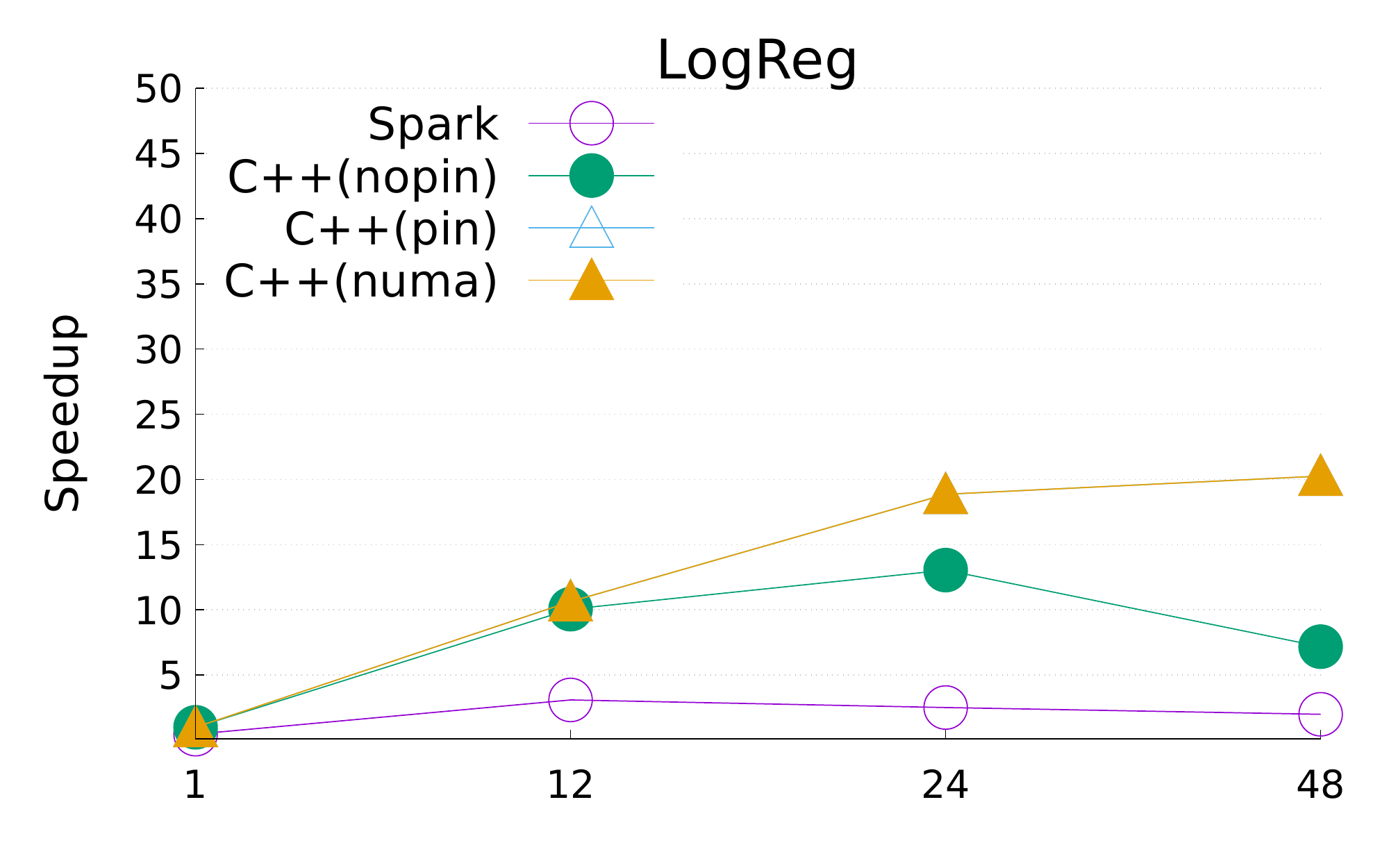}
	\hfill
	\includegraphics[width=0.49\linewidth]{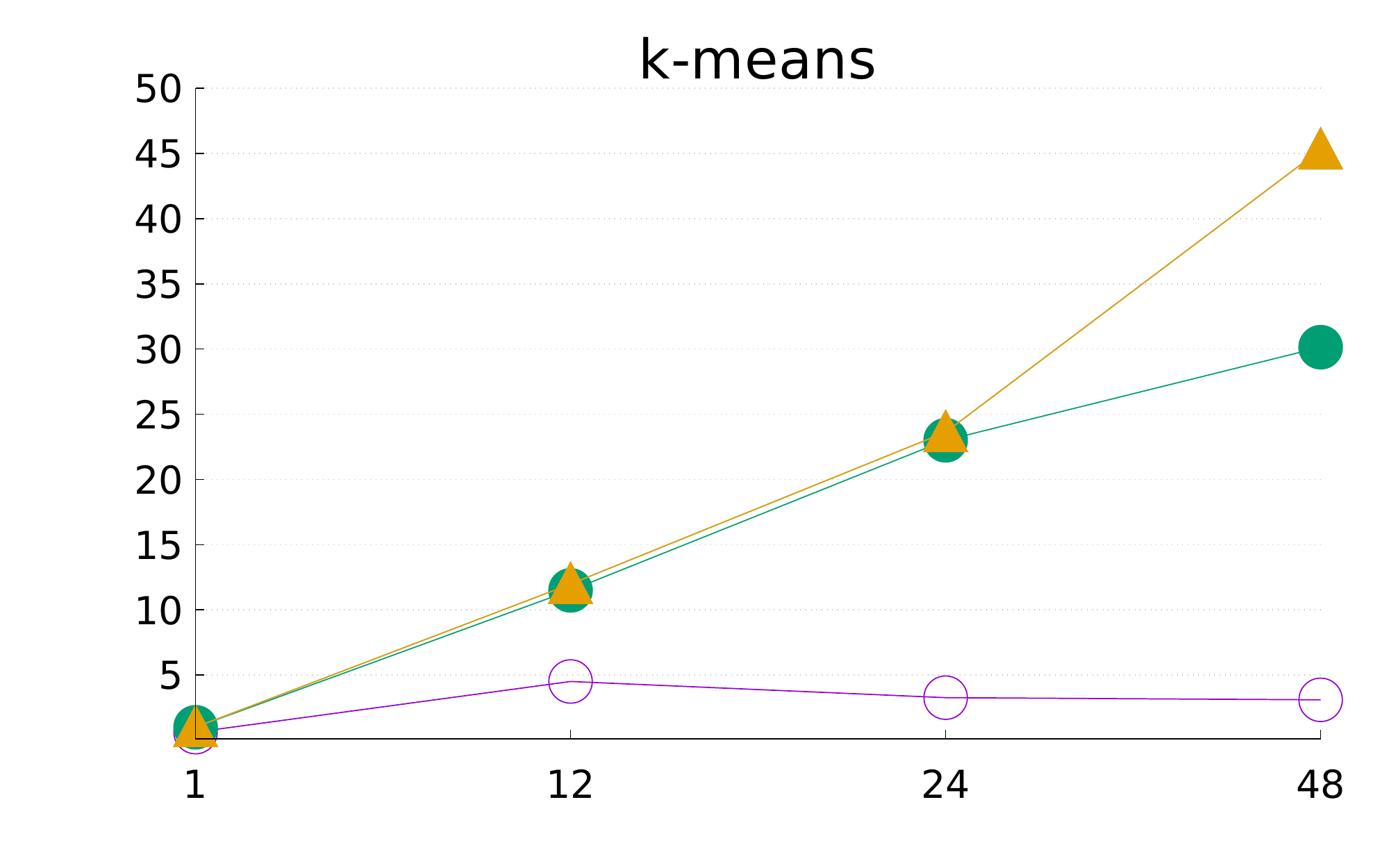}
	\\
	\includegraphics[width=0.49\linewidth]{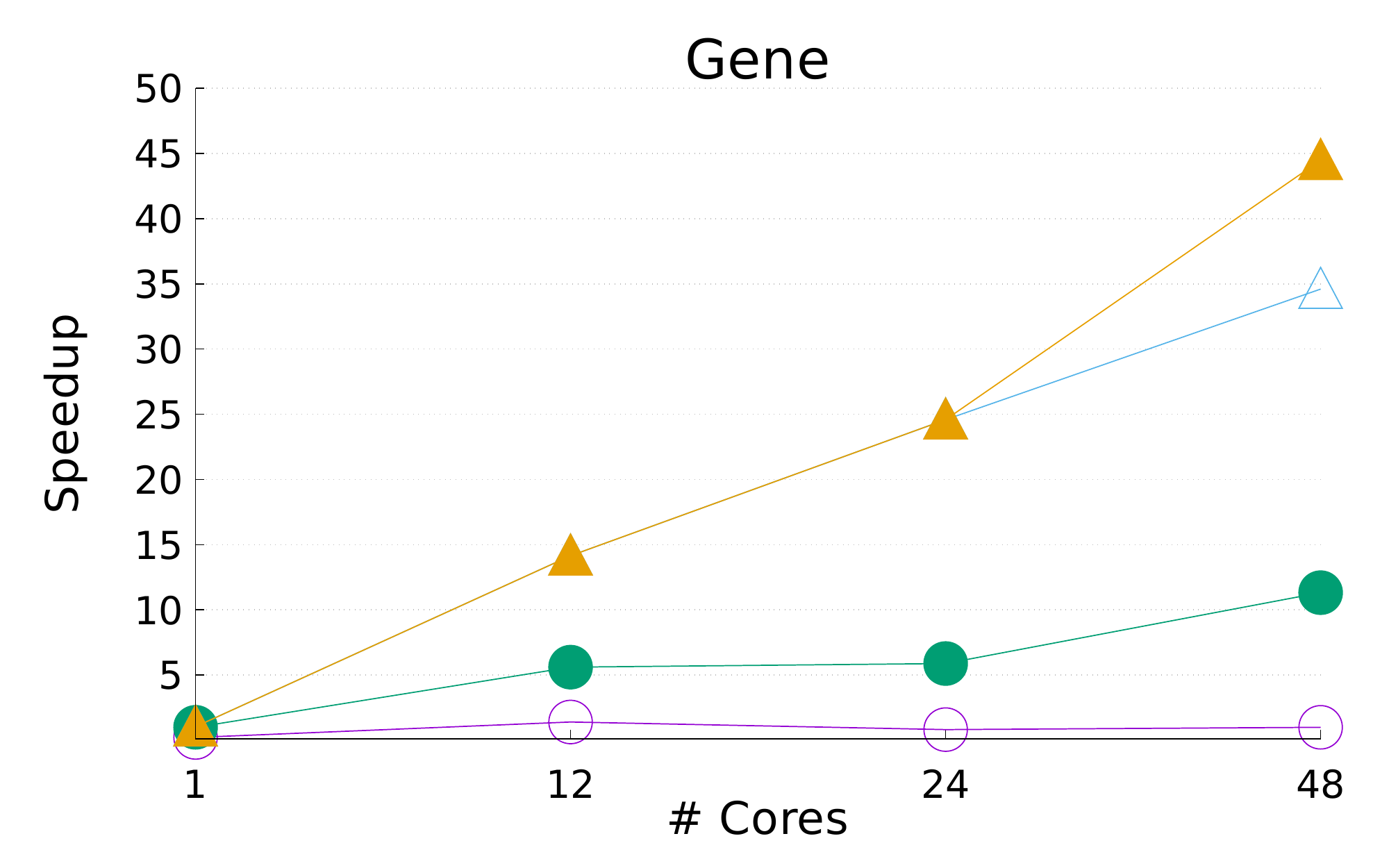}
	\hfill
	\includegraphics[width=0.49\linewidth]{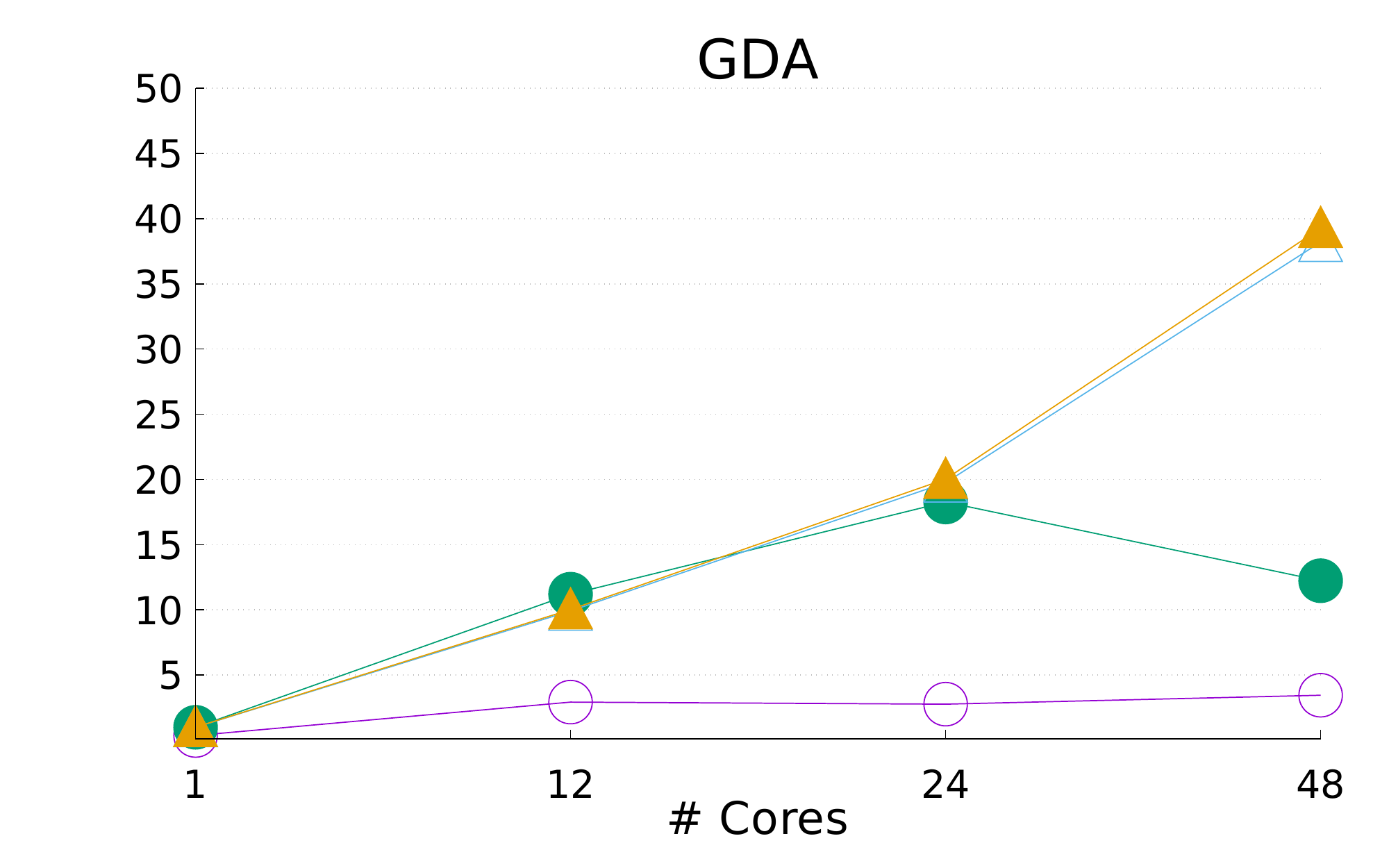}
	\caption{\label{fig:delite-numa}Level 3: Machine learning kernels, scaling
	on shared memory NUMA with thread pinning and data partitioning \vspace{-2ex}}
\end{figure}

\vspace{-2ex}
\paragraph*{Shared-Memory NUMA}
In a recent study, Brown et al.~\cite{Brown2016GCO} compared Spark with Delite with regards to NUMA scalability on a single shared-memory machine with 4 CPU sockets and a total of 48 cores. Specifically, Gaussian Discriminant Analysis (GDA), logistic regression (LogReg), k-means clustering, and a gene barcoding application (Gene) were chosen as machine learning benchmarks. As shown in Figure~\ref{fig:delite-numa}, Delite gains significant speedups over Spark in every application studied, with thread pinning and NUMA-aware data partitioning contributing in different ways for each application. As stated previously, Flare Level 3 generates code in Delite's intermediate language DMLL, and we have verified that the generated code matches perfectly the code in~\cite{Brown2016GCO}, thus guaranteeing the same performance.

\begin{figure}
	\centering

	\includegraphics[width=\linewidth]{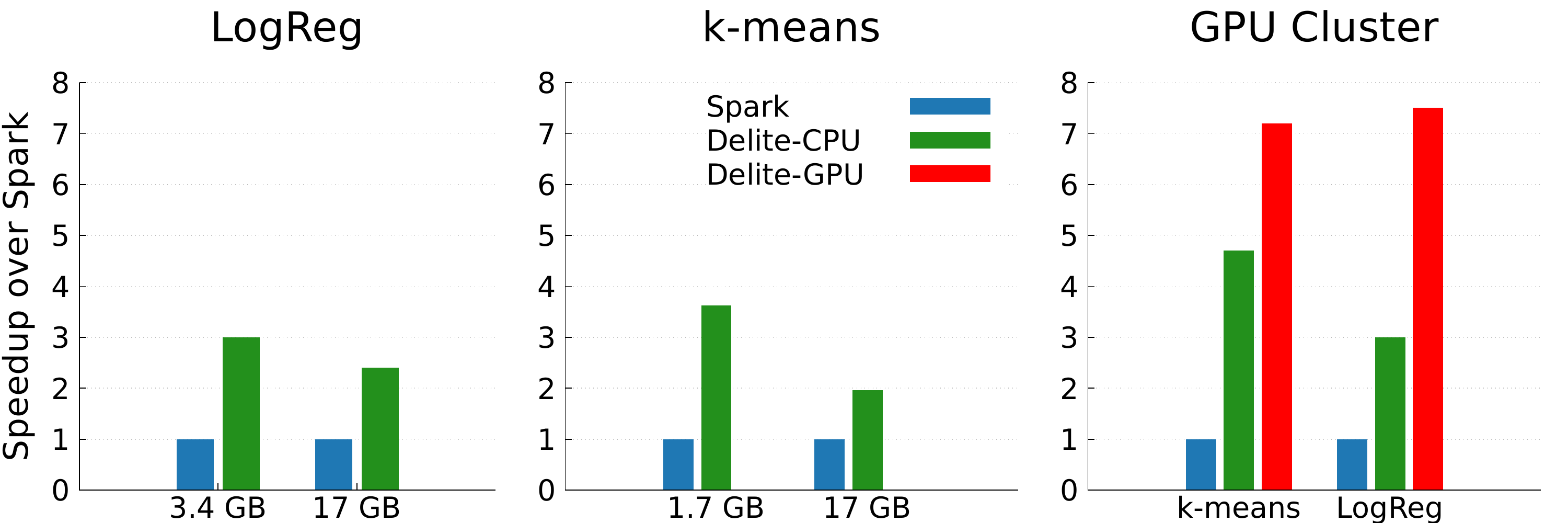}

	\caption{\label{fig:delite-cluster}Level 3: Machine learning kernels running on a 20 node Amazon cluster (left, center) and on a 4 node GPU cluster connected within a single rack.\vspace{-2ex}}

\end{figure}

\vspace{-2ex}
\paragraph*{Clusters and GPU}
Brown et al.~\cite{Brown2016GCO} also showcase speedups over Spark when running on a small cluster of machines, and when employing GPU accelerators on the k-means and LogReg applications. 
Despite running on Spark's intended architecture, when run on Amazon EC2 using 20 nodes, the code generated by Delite demonstrated a 2$\times$-3.5$\times$ speedup over Spark for k-means, and approximately a 2.5$\times$-3$\times$ speedup for LogReg (see Figure~\ref{fig:delite-cluster}, adapted from~\cite{Brown2016GCO}). When these applications were moved to a cluster of higher-end machines with more CPU cores as well as GPUs, this performance gap widened even further. The study shows that when running on a GPU cluster of 4 nodes, each with 12 cores, performance jumped to above a 7$\times$ speedup for each application. Again, for completeness, Flare Level 3 generates DMLL, and we verified that all generated code was identical to that used in~\cite{Brown2016GCO}.


\vspace{-2ex}
\section{Related Work}
\label{sec:related}

Numerous \textbf{cluster computing frameworks} implement a combination of parallel, distributed, relational, procedural, and MapReduce computations. The MapReduce model \cite{DBLP:conf/osdi/DeanG04} realized in Hadoop \cite{hadoop} performs big data analysis under shared-nothing, unreliable machines. Twister \cite{ekanayake2010twister} and Haloop \cite{bu2010haloop} support iterative MapReduce workloads by avoiding reading unnecessary data and keeping invariant data between iterations. Likewise, Spark \cite{zaharia10spark,SparkCACM} tackles the issue of data reuse among MapReduce jobs or applications by explicitly persisting intermediate results in memory. Along the same lines, the need for an expressive programming model to perform analytics on structured and semistructured data motivated Hive \cite{thusoo2009hive}, Dremel \cite{melnik2010dremel}, Impala \cite{bittorf2015impala}, Shark \cite{xin2013shark} and Spark SQL \cite {DBLP:conf/sigmod/ArmbrustXLHLBMK15} and many others. SnappyData \cite{DBLP:conf/sigmod/RamnarayanMWMKB16} integrates Spark with a transactional main-memory database to realize a unified engine that supports streaming, analytics and transactions. Moreover, Asterix \cite{behm2011asterix}, Stratosphere \cite{alexandrov2014stratosphere}, and Tupleware \cite{DBLP:journals/pvldb/CrottyGDKBCZ15} provide data flow-based programming models and support user defined functions (UDFs). Flare distinguishes itself by maintainng expressiveness while achieving performance close to highly optimized systems like HyPer \cite{DBLP:journals/pvldb/Neumann11}. Flare keeps Spark SQL's front-end intact, and integrates its own code generation and runtime (in Level 2) replacing Spark's RDDs and distributed-by-default runtime system. Furthermore, Flare Level 3 integrates with Delite as its execution back-end in order to support multi-DSL applications. By design, Flare's main target is scaled-up in-memory clusters where faults are improbable.

\textbf{Query Compilation} Recently, code generation for SQL queries 
has regained momentum. Historic efforts go back all the way to System R \cite{astrahan1976system}. Query compilation can be realized using code templates. e.g., Daytona \cite{greer1999daytona} or HIQUE \cite{krikellas2010generating}, general purpose compilers, e.g., HyPer \cite{DBLP:journals/pvldb/Neumann11} and Hekaton \cite{diaconu2013hekaton}, or DSL compiler frameworks, e.g., Legobase \cite{klonatos2014building}, DryadLINQ \cite{yu2008dryadlinq}, and DBLAB \cite{DBLP:conf/sigmod/ShaikhhaKPBD016}.

\textbf{Embedded DSL frameworks and intermediate languages} address the compromise between productivity and performance in writing programs that can run under diverse programming models. 
Voodoo \cite{pirk2016voodoo} addresses compiling portable query plans that can run on CPUs and GPUs. Voodoo's intermediate algebra is expressive and captures hardware optimizations, e.g., multicores, SIMD, etc. Furthermore, Voodoo is used as an alternative back-end for MonetDB \cite{boncz2005monetdb}. Delite \cite{sujeeth2014delite}, a general purpose compiler framework, implements high-performance DSLs (e.g., SQL, Machine Learning, graphs and matrices), provides parallel patterns and generates code for heterogeneous targets. The Distributed Multiloop Language (DMLL)  \cite{Brown2016GCO} provide rich collections and parallel patterns and supports big-memory NUMA machines. Weld \cite{palkarweld} is another recent system that aims to provide a common runtime for diverse libraries e.g., SQL and machine learning. Steno \cite{murray2011steno} performs optimizations similar to DMLL to compile LINQ queries. Furthermore, Steno uses DryadLINQ \cite{yu2008dryadlinq} runtime for distributed execution. Nagel et. al. \cite{nagel2014code} generates efficient code for LINQ queries. Weld  is similar to DMLL in supporting nested parallel structures. However, Weld focuses on a larger set of data science frameworks (Spark, TensorFlow, NumPy and Pandas) and does not, e.g, support a large enough set of joins and other operators to run all TPC-H queries.
Query compilation inside Flare Level 3 is based on Delite's OptiQL DSL and DMLL intermediate language.

\textbf{Performance evaluation} in data analytics frameworks aims to identify performance bottlenecks and study the parameters that impact performance the most, e.g., workload, scaling-up/scaling-out resources, probability of faults, etc. A recent study \cite{DBLP:conf/nsdi/OusterhoutRRSC15} on a single Spark cluster revealed that CPU, not I/O, is the source of bottlenecks. McSherry et al.~\cite{DBLP:conf/hotos/McSherryIM15} proposed the COST (Configuration that Outperforms a Single Thread) metric, and showed that in many cases, single-threaded programs can outperform big data processing frameworks running on large clusters. TPC-H \cite{tpchbib} is a decision support benchmark that consists of 22 analytical queries that address several ``choke points,'' e.g., aggregates, large joins, arithmetic computations, etc. \cite{boncz2013tpc}. Flare Level 2's performance evaluation is done using TPC-H.

\section{Conclusion}
\label{sec:conclusions}
Modern data analytics need to combine multiple programming models and make efficient use of modern hardware with large memory, many cores, and accelerators such as GPUs. We introduce Flare: a new back-end for Spark that brings relational performance on par with the best SQL engines, and also enables highly optimized of heterogeneous workloads. Most importantly, all of this comes without giving up the  expressiveness of Spark's high-level APIs. 
\begin{sloppypar}
We believe that multi-stage APIs, in the spirit of DataFrames, and compiler systems like Flare and Delite, will play an increasingly important role in the future to satisfy the increasing demand for flexible and unified analytics with high efficiency.
\end{sloppypar}

\bibliographystyle{abbrv}
\bibliography{ppl,references,refs}

\end{document}